\documentclass[final,3p,times]{elsarticle}

\usepackage{amssymb,amsmath,amsfonts}
\usepackage{tabularx}
\usepackage{xcolor}

\usepackage{algorithm}
\usepackage[boxruled,algo2e]{algorithm2e}
\usepackage[hidelinks]{hyperref}
\DeclareMathOperator*{\argmin}{\arg\!\min}
\newcommand\norm[1]{\left\lVert#1\right\rVert}

\usepackage{hyperref}

\journal{Journal of Computational Physics}

\begin{document}

\begin{frontmatter}

\title{Jacobian-Scaled K-means Clustering for Physics-Informed Segmentation of Reacting Flows}

\author[inst1,inst2]{Shivam Barwey\corref{cor1}}
\author[inst1]{Venkat Raman}
\cortext[cor1]{Corresponding author. E-mail address: \href{mailto:sbarwey@anl.gov}{sbarwey@anl.gov} (S. Barwey)}

\affiliation[inst1]{organization={Department of Aerospace Engineering, University of Michigan},
            addressline={1320 Beal Ave.}, 
            city={Ann Arbor},
            postcode={48109}, 
            state={MI},
            country={USA}}
\affiliation[inst2]{organization={Argonne Leadership Computing Facility, Argonne National Laboratory},
            addressline={9700 S Cass Ave.}, 
            city={Lemont},
            postcode={60439}, 
            state={IL},
            country={USA}}

\begin{abstract}
This work introduces Jacobian-scaled K-means (JSK-means) clustering, which is a physics-informed clustering strategy centered on the K-means framework. The method allows for the injection of underlying physical knowledge into the clustering procedure through a distance function modification: instead of leveraging conventional Euclidean distance vectors, the JSK-means procedure operates on distance vectors scaled by matrices obtained from dynamical system Jacobians evaluated at the cluster centroids. The goal of this work is to show how the JSK-means algorithm -- without modifying the input dataset -- produces clusters that capture regions of dynamical similarity, in that the clusters are redistributed towards high-sensitivity regions in phase space and are described by similarity in the source terms of samples instead of the samples themselves. The algorithm is demonstrated on a complex reacting flow simulation dataset (a channel detonation configuration), where the dynamics in the thermochemical composition space are known through the highly nonlinear and stiff Arrhenius-based chemical source terms. Interpretations of cluster partitions in both physical space and composition space reveal how JSK-means shifts clusters produced by standard K-means towards regions of high chemical sensitivity (e.g., towards regions of peak heat release rate near the detonation reaction zone). The findings presented here illustrate the benefits of utilizing Jacobian-scaled distances in clustering techniques, and the JSK-means method in particular displays promising potential for improving former partition-based modeling strategies in reacting flow (and other multi-physics) applications.
\end{abstract}


\end{frontmatter}

\section{Introduction}
\label{sec:intro}
Accelerating computational tools for simulating complex reacting flows will open new pathways for design and optimization of propulsion and energy conversion systems \cite{raman2019emerging,montans2019data}. One approach is the use of data-driven models that can utilize high-fidelity simulations or experimental data to generate a new class of robust and fast computational tools. For instance, data-based models are typically trained on a large collection of high-fidelity flow field snapshots derived from these big data sources. The general goal is to use these models to effectively replace highly expensive physics-based CFD solvers with either surrogates that replace the entire solver, or acceleration modules that replace only the computationally prohibitive components of the solver. For example, in the surrogate modeling strategy, time-resolved data may be used to train prognostic models in low-dimensional latent spaces produced by modal decompositions \cite{mezic_modal_decomp,shivam_ctm} or neural network (NN) based autoencoders \cite{carlberg_2020,romit_rnn_2021}. In the alternate strategy, computationally prohibitive subroutines can be replaced using methods like learned field transformations \cite{bunny_paper}, flow field super-resolution \cite{kim_super_res_2021}, and NN-based stiff time integrators \cite{masri_ann,neuralode_stiff}. Ultimately, when proper care is taken during the training stage, these data-based models have the powerful ability to capture highly complex physics contained in the training data, leading to models that can outperform purely physics-based counterparts in terms of faster evaluation times, improved prediction accuracy, or both. 

Despite their advantages and demonstrated success, purely data-based approaches have limitations. The principal disadvantage, alongside the potentially time-consuming training stage, is the tendency for these models to overfit to specific geometries, flow regimes/configurations, and boundary conditions. These issues are especially problematic when attempting to produce data-based models of highly nonlinear systems characterized by multi-scale and multi-physics fluid flow behavior in complex geometries. For example, in energy generation and hypersonic propulsion applications, the macroscopic behavior of systems like stationary gas turbines, detonation engines, and scramjets, are described by turbulence-chemistry interactions at both low and high Mach numbers. Data-based predictions for the highly nonlinear and unsteady reacting flow processes that arise at these operating conditions can become unreliable in extrapolative settings, leading to their limited usage in engineering design optimization and downstream modeling workflows \cite{raman2019emerging}. 

To address these limitations, state-of-the-art model development for multi-physics systems has centered on injecting physical intuition into purely data-based strategies. As a result, the parameters in data-based models become constrained to some degree by the underlying partial differential equations (PDEs), which are the Navier-Stokes equations in the case of fluid flows. This need for physical consistency in data-based frameworks has led to significant research into physics-constrained (or physics-guided) data-driven modeling approaches based on various types of NNs, for which fast, GPU-friendly optimization algorithms have been matured by the data science and machine learning (ML) communities \cite{karniadakis2021physics}. Modern examples of such strategies include (a) physics-informed neural networks (PINNs) \cite{raissi_pinn}, which leverage automatic differentiation to cast the objective function as the governing PDEs, (b) geometric deep learning \cite{geometric_dl} and graph neural networks \cite{chamberlain2021grand}, which utilize neighborhood aggregation strategies consistent with flow invariants and numerical discretization schemes, and (c) statistically-constrained generative modeling frameworks \cite{wu_2020,malik_gan,tony_gan}, which ensure predictions are consistent with macroscopic physical trends.

In the rapidly developing field of physics-informed data-based modeling, much less attention has been paid to unsupervised methods based on data clustering (with some notable exceptions \cite{physics_clustering}). Embedding physical knowledge into unsupervised clustering objectives can produce valuable acceleration pathways that offer unique advantages for multi-physics model development. To this end, the objective of this work is to develop a physics-informed clustering strategy based on the K-means framework \cite{kmeans_main_reference}, termed Jacobian-scaled K-means (JSK-means) clustering. In this work, alongside presenting the relevant methodology, the new method is applied to produce improved, physically-consistent segmentations of compressible reacting flow fields found in high-speed detonation-based propulsion applications.

To motivate the need for the JSK-means approach, it is important to first highlight the concept of localized modeling, which involves generating partitions of the phase space (or feature space) in which the relevant data samples reside. The resulting partitions are subsequently used as conditioning variables for downstream modeling tasks, leading to a type of classification-guided regression workflow. Ideally, if the discovered partitions (say, generated by a clustering algorithm) are able to isolate different physical regimes of interest, the conditioning process leads to both accurate and interpretable predictive capability that can be leveraged for multi-physics modeling. As a result, these strategies in recent years have been successfully developed for many complex fluid dynamics applications -- examples include generating reduced-order-models via clustering over flow field snapshots \cite{kaiser_crom,gunzburger_cvt_rom}, creating multi-regime wall models in turbulent flows via self-organizing maps \cite{meneveau_som}, and constructing closure models for dynamical systems using conditional averages over trajectories \cite{pope_ten,shivam_timeaxis}. 

In reacting flows, which is the application context of this work, localized models have been almost exclusively used to accelerate detailed chemical kinetics routines (particularly the evaluation of the species source terms), as these are the most time-consuming components in high-fidelity reacting CFD simulations \cite{jacki_petascale,detfoam_paper}. These methods have been shown to produce more accurate predictions of unsteady flow phenomena when compared to global counterparts (e.g. proper orthogonal decomposition) \cite{shivam_ftc,dalessio_2020}. Within this context, examples of localized models include (a) physics-based in-situ adaptive approaches \cite{pope_isat,prism,liang2015pre}, in which tabulations of the state space are obtained from chemical source term characterizations, and (b) data-based partitioning approaches, in which partitions based on unsupervised algorithms like local principal component analysis \cite{parente_mgpca}, random forests \cite{lei_random_forest}, and K-means clustering \cite{shivam_ftc}, are used to delineate combustion regimes and deploy targeted models for otherwise expensive chemical source term evaluations. 

The above methods rely on a notion of similarity between two points in feature space to produce the partitions or clusters. In most cases, this similarity comes from a standard Euclidean distance measure. In reaction-dominated flows, the driving features correspond to species mass fractions and temperature. As such, distance measures in these applications encode deviations in the thermochemical composition space. The goal of this work is to augment the notion of Euclidean distance with knowledge of the underlying governing equations, such that the distance between two points in feature space reflects not only feature similarity, but also \textit{dynamical similarity}. This concept is applied here to the K-means clustering framework, resulting in the aforementioned variant termed Jacobian-Scaled K-means (JSK-means) clustering. This work shows, using a target reacting flow dataset sourced from complex channel detonation simulations, how such a modification to the distance function effectively pushes clusters towards regions in thermochemical composition space that reduce within-cluster variation in chemical source term instead of composition, resulting in a novel pathway for physics-informed localized modeling. It is emphasized that, although applied here to a reacting flow dataset, the methodology presented in this work incorporates a general distance function modification that can be used to produce dynamically-consistent partitions of feature spaces in many other physics-based applications.

\section{Dataset}
\label{sec:dataset}
This section describes the simulation procedure used to generate the clustering dataset. More specifically, details on the simulation configuration, flow solver numerics, and pre-processing steps are provided. 

\begin{figure}
    \centering
    \includegraphics[width=0.6\columnwidth]{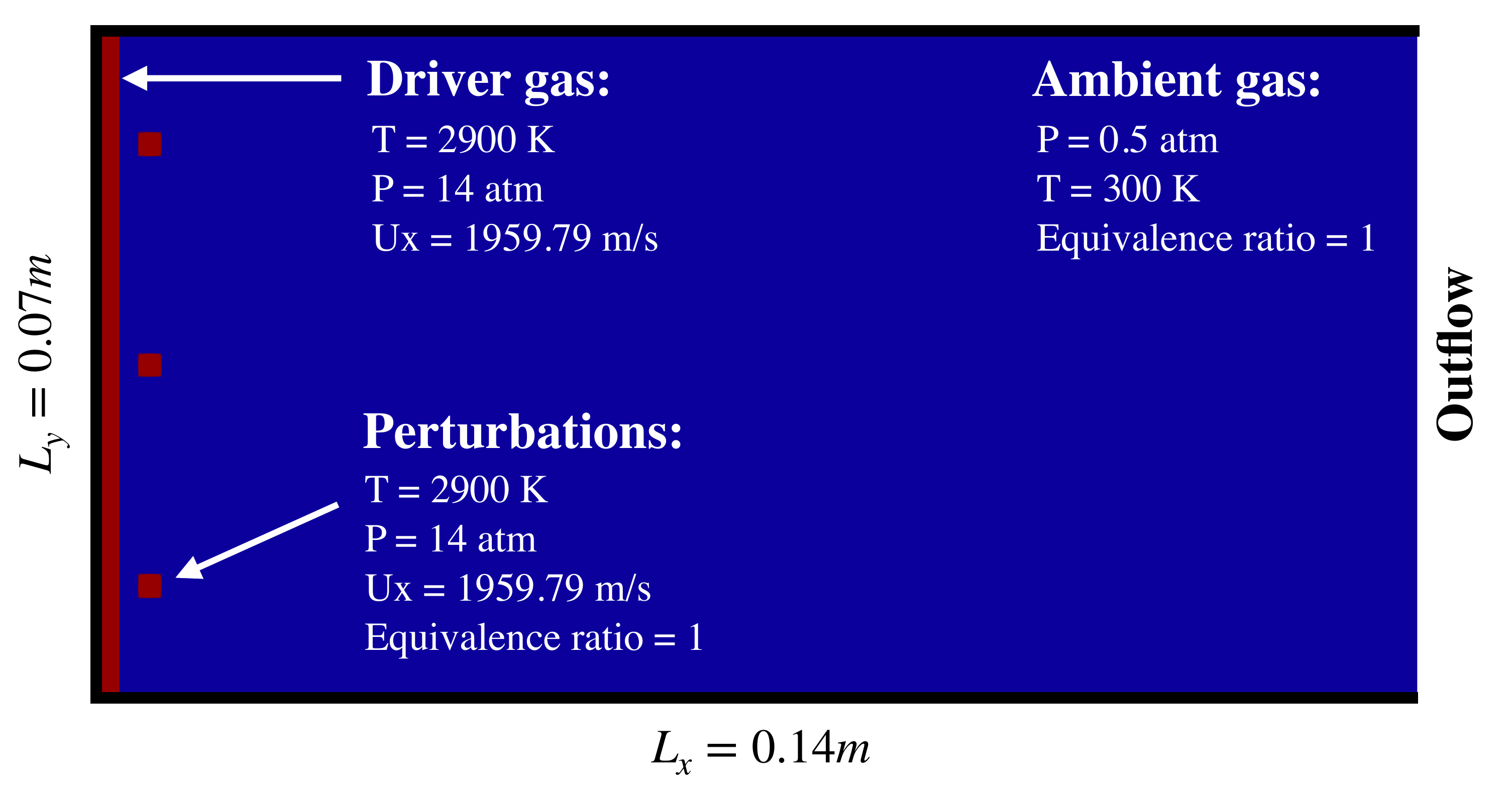}
    \caption{Computational domain for channel detonation simulation.}
    \label{fig:configuration}
\end{figure}

The dataset used in this analysis comes from high-fidelity hydrogen-air detonation simulations in a 2-dimensional channel configuration, as shown in Fig.~\ref{fig:configuration}. The flow solver is implemented with the finite-volume method in the AMReX library \cite{amrex}, and the governing equations are the compressible reacting Navier-Stokes equations. Although the solver is implemented in the AMReX framework, the simulations conducted for this study do not utilize adaptive mesh refinement (all data were collected from single-level runs). The numerics are globally 2nd-order and are consistent with those used in UMReactingFlow \cite{detfoam_paper}. Chemical kinetics for hydrogen-air combustion is described by the mechanism of Mueller et. al \cite{hon_mueller}, and consists of 9 species ($\text{H}_2, \text{O}_2, \text{H}_2\text{O}, \text{H}, \text{O}, \text{OH}, \text{HO}_2, \text{H}_2 \text{O}_2, \text{N}_2$) and 21 reactions. Equipped with this mechanism, the open-source library Cantera is used to evaluate chemical source terms, transport coefficients, and other relevant thermodynamic quantities \cite{cantera}.

As shown in Fig.~\ref{fig:configuration}, the channel is $0.07$~m in the y-direction and $0.14$~m in the x-direction. The boundary conditions consist of slip walls everywhere except for the rightmost boundary, which is an outflow. The unsteady detonation is initiated by prescribing a high-energy driver gas in a $2$~mm region on the leftmost side of the domain. The temperature, pressure, and x-component of the velocity for the driver gas is set to $2900$~K, $14$~atm, and the Chapman-Jouguet (CJ) speed, respectively (the y-component of the velocity is zero everywhere initially). The species mass fractions for the driver gas are obtained from the CJ condition evaluated from the idealized Zeldovich-von Neumann-D\"{o}ring (ZND) detonation solution (the Caltech Shock and Detonation toolbox was used to this end \cite{sdtoolbox}). Similar to the approach used in Ref.~\cite{supraj_aiaa}, roughly $2$~mm ahead of the driver gas, an array of cube-like perturbations of high-energy stoichiometric hydrogen-air mixture is placed roughly $2$~mm ahead of the driver gas region. These perturbations are required to initiate transverse wave reflections that produce the characteristic triple point structures observed in unsteady detonations in 2 and 3 dimensions. The composition for the ambient gas is stoichiometric H2/Air at $0.5$~atm and $300$~K.

The domain is discretized with 4096 cells in the x-direction and 2048 cells in the y-direction, producing a total cell count of roughly 8 million and a grid resolution of $\Delta x \approx 34$ micron. At an ambient pressure of $0.5$~atm, this grid resolution results in roughly $10$ cells within the ZND induction zone and therefore exceeds required spatial resolution limit for resolved unsteady H2/Air detonations \cite{supraj_stratification}. Starting from the conditions specified in Fig.~\ref{fig:configuration}, the simulation was run for a total time of $10^{-4}$ seconds at a Courant-Friedrichs-Lewy (CFL) number of 0.6, which was enough time to ensure progression of the developed CJ detonation through the entirety of the channel.

A subset of the full channel domain at $t=50$~$\mu$s that contains the entirety of the detonation wave structure is extracted in a cropping procedure. The cropped subdomain (bounded by $x=[0.085,0.95]$ and $y=[0,0.07]$ meters) produces a set of $N=598015$ thermochemical state vector samples that describe all relevant details of the detonation wave structure (e.g. triple point oscillations, transverse waves, deflagration regions, ambient regions, etc.) and ignores the regions beyond the sonic choke point that are irrelevant to the wave dynamics. 

\begin{figure}
    \centering
    \includegraphics[width=\columnwidth]{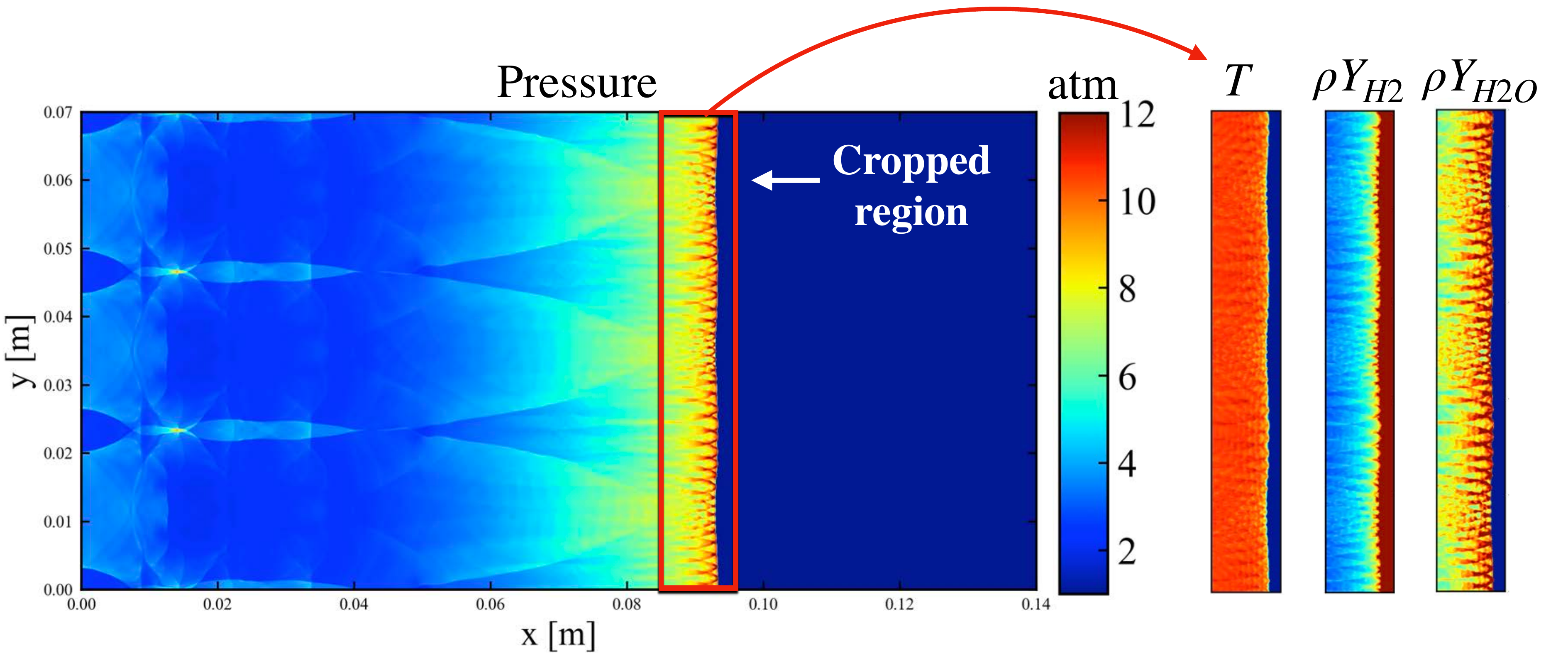}
    \caption{Cropping procedure for an instantaneous detonation flowfield at $t=50$~$\mu$s used to generate the clustering dataset. Colorbar ranges for $T$, $\rho Y_{H2}$, and $\rho Y_{H2O}$ are $[300,3200]$~K, $[0.001,0.006]$~$\text{kg}/\text{m}^3\text{s}$, and $[0.1,0.2]$~$\text{kg}/\text{m}^3\text{s}$ respectively.}
    \label{fig:snapshot}
\end{figure}

The individual samples are denoted $\phi_i \in \mathbb{R}^D$, where $i=1,\ldots,N$. The $\phi_i$ reside in the 10-dimensional composition space defined by temperature and species mass concentrations (i.e. $D=10$). More specifically, $\phi_i = [T_i, \rho Y_{1,i}, \ldots, \rho Y_{{N_S},i}]^\text{T}$, where $T_i$ is the temperature for sample $i$, $\rho Y_k$ is the $k$-th species concentration, and T denotes the transpose operation (not to be confused with temperature). Through this composition vector, the local fluid density can be obtained from a summation over the species mass concentrations, and the pressure can be extracted from the ideal gas law. Using these composition space samples, two datasets are created: the first is the composition data matrix $\Phi = [\phi_1, \phi_2, \ldots, \phi_N] \in \mathbb{R}^{D \times N}$ and the second is the ground-truth chemical source term data matrix $\Omega = [S(\phi_1), S(\phi_2), \ldots, S(\phi_N)] \in \mathbb{R}^{D \times N}$, where $S_i(\phi) = [\frac{d T_i}{dt}, \frac{d \rho Y_{1,i}}{dt}, \ldots, \frac{d \rho Y_{{N_S},i}}{dt}]^\text{T}$ is referred to as the chemical source term vector for sample $i$. As per detailed finite-rate chemical kinetics standards, the species concentration source terms are evaluated using linear combinations of elementary Arrhenius-based reactions \cite{poinsot_book}. The temperature source term is evaluated under the assumption of a constant-volume reactor as
\begin{equation}
    \label{eq:temperature_ode}
    \frac{d T_i}{dt} = -\frac{1}{c_v} \sum_{k=1}^{N_S} \frac{\epsilon_k}{W_k} \frac{d \rho Y_k}{dt}, 
\end{equation}
where $c_v$ denotes the mass-based specific heat at constant volume for the mixture, $\epsilon_k$ is the molar internal energy for species $k$ (derived from NASA polynomials), $W_k$ is the species molecular weight, and $\frac{d \rho Y_k}{dt}$ is the source term for species concentration $k$. Ultimately, the composition data in $\Phi$ is used in the clustering procedure in Sec.~\ref{sec:methodology}, and the data in $\Omega$ is used to further assess clustering outputs.

\section{Methodology}
\label{sec:methodology}
\subsection{Standard K-means Clustering} 
\label{sec:kmeans}
Before describing the Jacobian-Scaled K-means clustering strategy in Sec.~\ref{sec:jskmeans}, a brief overview of the standard K-means algorithm is first presented. The reader is directed to Ref.~\cite{kmeans_main_reference} for a more detailed description of the baseline K-means algorithm. 

Given the thermochemical state data in $\Phi$, the standard K-means procedure minimizes the objective function 
\begin{equation}
    \label{eq:standard_kmeans_obj} 
    E_{\phi} = \sum_{k=1}^{K} \sum_{i=1}^{N} 
    \big \lVert 
    {\bf L}_{i,k} [ \phi_i - c_k ]
    \big \rVert_2^2
\end{equation}
based on the initialization and convergence of a finite set of $K$ centroids. The quantity $K$ is also equivalent to the number of clusters. The number of clusters $K$ is one of two major inputs to the K-means algorithm, the other being the initial locations of the centroids in the $D$-dimensional composition space. Equation~\ref{eq:standard_kmeans_obj} is interpreted as the total within-cluster variation of composition samples, and depends entirely on the centroid values. 

As the name implies, the centroid $c_k$ is defined as the arithmetic mean of all samples within the $k$-th cluster as
\begin{equation}
    \label{eq:standard_centroid} 
    c_k = \frac {\sum_{i=1}^{N} {\bf L}_{i,k} \phi_i } { \sum_{i=1}^{N} {\bf L}_{i,k} }.
\end{equation}
In both Eqs.~\ref{eq:standard_kmeans_obj} and \ref{eq:standard_centroid}, the quantity ${\bf L}_{i,k}$ denotes an integer value extracted from the $i$-th row and $k$-th column of the cluster assignment matrix  ${\bf L} \in \mathbb{Z}^{N \times K}$. This value in the cluster assignment matrix is unity if the respective sample is closest to $c_k$ in the Euclidean sense and zero otherwise. This is formally expressed as 
\begin{equation}
    \label{eq:standard_indicator}
    {\bf L}_{i,k} =     
    \begin{cases}
      1 & \text{if}\ k = \argmin\limits_{j}  \big \lVert \phi_i - c_j  \big \rVert_2^2, \ \text{and} \\
      0 & \text{otherwise}.
    \end{cases}
\end{equation}
The standard K-means algorithm (the well-known Lloyd's algorithm) consists of alternating between centroid updates via Eq.~\ref{eq:standard_centroid} and cluster label updates via Eq.~\ref{eq:standard_indicator} until the centroid positions in $\mathbb{R}^D$ change by a negligibly small amount. This procedure is described in Algorithm~\ref{alg:standard_kmeans}.

\begin{algorithm}
\caption{K-means Clustering (Lloyd's Algorithm)}
\label{alg:standard_kmeans}
\KwData{ \\ 
    (1) Set number of clusters $K$\\
    (2) Set convergence error tolerance $\varepsilon_{tol}$ \\
    (3) Initialize dataset $\Phi$ to be clustered \\ 
    (4) Initialize centroids: ${\bf c}_{new} \gets [c_1, \ldots, c_K]$ \\ 
    (5) Initialize convergence criterion: $\varepsilon_{c} \gets \text{Inf}$ \\
}
\vspace{0.5\baselineskip}
\KwResult{ \\ 
    (1) Converged set of centroids in ${\bf c}_{new}$. \\
    (2) Cluster assignment matrix ${\bf L}$ (Eq.~\ref{eq:standard_indicator}) \\
}
\vspace{0.5\baselineskip}

\While{$\varepsilon_{c} \geq \varepsilon_{tol}$} 
{
    a) Copy centroids: ${\bf c}_{old} \gets {\bf c}_{new}$ \\ 
    b) Update cluster assignment matrix ${\bf L}$ using ${\bf c}_{old}$ (Eq.~\ref{eq:standard_indicator})\\
    d) Update centroids in ${\bf c}_{new}$ using cluster assignment matrix ${\bf L}$ (Eq.~\ref{eq:standard_centroid}) \\ 
    e) Compute convergence criterion: $\varepsilon_c \gets \lVert {\bf c}_{new} - {\bf c}_{old} \rVert_F$  \\ 
    f) Break if specified number of iterations is reached
}
\end{algorithm}

Centroid convergence rates are highly dependent on initial centroid locations. The naive approach is to simply initialize centroids randomly -- this is ill-advised, because random initialization not only scales poorly with increase in sample dimensionality $D$, but also faces the risk of initializing two or more centroids close to one another in the phase space, which increases the chances of producing redundant clusters and sub-optimal values for the objective function. The K-means++ is an alternative initialization procedure that addresses these issues \cite{kmeanspp}. Instead of picking the initial locations arbitrarily, K-means++ initializes centroids with the goal of achieving a large amount of separation between the centroids. However, the K-means++ initialization is stochastic, whereas the K-means algorithm itself is deterministic for a fixed initial set of centroids. Therefore, convergence of Alg.~\ref{alg:standard_kmeans} using multiple realizations of starting centroid locations is necessary to ensure statistical confidence in the K-means output \cite{shivam_ctm}.

\subsection{Jacobian-Scaled K-means Clustering} 
\label{sec:jskmeans}
Instead of minimizing the standard K-means objective described Eq.~\ref{eq:standard_kmeans_obj}, the goal of the Jacobian-Scaled K-means (JSK-means) clustering strategy is to minimize
\begin{equation}
    \label{eq:jacobian_kmeans_obj} 
    E_{{\bf A}} = \sum_{k=1}^{K} \sum_{i=1}^{N} 
    \big \lVert 
    {\bf L}^{{\bf A}}_{i,k}  {\bf A}_k  [\phi_i - c_k]
    \big \rVert_2^2.
\end{equation}

There are two main differences between Eq.~\ref{eq:jacobian_kmeans_obj} and the standard K-means objective in Eq.~\ref{eq:standard_kmeans_obj}. The first difference is the presence of the weighting matrix ${\bf A}_{k} \in \mathbb{R}^{D \times D}$, which acts as a linear transformation, or scaling, on the sample-to-centroid distance vector $\phi_i - c_k$. As denoted by the subscript $k$, this matrix is cluster-dependent through dependency on the centroid (i.e. ${\bf A}_{k} = {\bf A}(c_k)$). As such, the objective in Eq.~\ref{eq:jacobian_kmeans_obj} comes from a modification to the standard Euclidean distance function used to define sample similarity in the phase space. More specifically, if the Euclidean sample-centroid distance function is given as 
\begin{equation}
    \label{eq:standard_distance}
    d(\phi_i, c_k) = \lVert \phi_i - c_k \rVert _2^2, 
\end{equation}
the modified distance function implied by Eq.~\ref{eq:jacobian_kmeans_obj} is 
\begin{equation}
    \label{eq:jacobian_distance} 
    d(\phi_i, c_k) = \lVert {\bf A}_k (\phi_i - c_k) \rVert _2^2.
\end{equation}
The second difference from the standard K-means approach is a modified assignment matrix, ${\bf L}^{{\bf A}}$. The cluster assignment mechanism in JSK-means is based on the modified distance measure in Eq.~\ref{eq:jacobian_distance} as 
\begin{equation}
    \label{eq:jacobian_indicator}
    {\bf L}^{{\bf A}}_{i,k} =     
    \begin{cases}
      1 & \text{if}\ k = \argmin\limits_{j}  \big \lVert {\bf A}_j (\phi_i - c_j)  \big \rVert_2^2, \ \text{and} \\
      0 & \text{otherwise}.
    \end{cases}
\end{equation}
Note that although the above cluster assignment matrix ${\bf L}^{\bf A}$ provides the same functionality as the standard K-means counterpart in Eq.~\ref{eq:standard_indicator} (i.e., it identifies the samples belonging to a particular cluster), the distance function used to prescribes the sample-centroid assignment has changed. The centroid $c_k$ in the JSK-means approach is defined in the same way as in the standard approach (Eq.~\ref{eq:standard_centroid}), but instead uses this modified cluster assignment matrix: 
\begin{equation}
    \label{eq:jacobian_centroid} 
    c_k = \frac {\sum_{i=1}^{N} {\bf L}^{{\bf A}}_{i,k} \phi_i } { \sum_{i=1}^{N} {\bf L}^{{\bf A}}_{i,k} }.
\end{equation}
The embedding of physical knowledge comes from the definition of the cluster-dependent weighting matrix ${\bf A}_k$. It is assumed that there exists a set of governing equations that describes the evolution of the phase space vector $\phi$ as 
\begin{equation}
    \label{eq:phase_space_dynamics} 
    \frac{d \phi}{dt} = {S}(\phi) \in \mathbb{R}^D, 
\end{equation}
where the quantity ${S}(\phi)$ is a velocity in $\mathbb{R}^D$. As described in Sec.~\ref{sec:dataset}, $\phi$ here is the thermochemical state vector. As such, the phase space velocity is the chemical source term, and the ordinary differential equation in Eq.~\ref{eq:phase_space_dynamics} describes the evolution of species mass fraction and temperature in composition space. 

The method for obtaining ${\bf A}_k$ is now described. The leading-order Taylor expansion for the chemical source term via Eq.~\ref{eq:phase_space_dynamics} is 
\begin{equation}
    \label{eq:taylor_expansion}
    {S}(\phi^q) - {S}(\phi^{ref}) =  
    \left.\frac{\partial {S}(\phi)}{\partial \phi} \right\rvert_{\phi=\phi^{ref}} \left( \phi^q - \phi^{ref} \right) + {\cal O}(\lvert\delta \phi\rvert^2), 
\end{equation}
where $\phi^{ref}$ is a reference point in composition space, $\phi^q$ is a query point, and $\frac{\partial {S}(\phi)}{\partial \phi}$ is the chemical Jacobian matrix. By casting the reference point $\phi^{ref}$ as the centroid $c_k$, the the expression for the scaling matrix ${\bf A}_k$ that acts on the sample-to-centroid distance vector emerges as the chemical Jacobian evaluated at the centroid:
\begin{equation}
    \label{eq:chemical_jacobian} 
    {\bf A}_k = \left.\frac{\partial {S}(\phi)}{\partial \phi} \right\rvert_{\phi=c_k}.
\end{equation}

By means of the Taylor expansion in Eq.~\ref{eq:taylor_expansion} and the definition of the scaling matrix in Eq.~\ref{eq:jacobian_centroid}, the distance function used in the JSK-means approach (Eq.~\ref{eq:jacobian_distance}) approximates the source term distance. More formally,
\begin{equation}
    d(\phi_i, \phi_j) = \lVert {\bf A}_k (\phi_i - \phi_j) \rVert _2^2 \approx \lVert {S}(\phi_i) - {S}(\phi_j) \rVert_2^2.
\end{equation}

Ultimately, by representing the scaling matrix as the chemical Jacobian, the objective function for JSK-means in Eq.~\ref{eq:jacobian_kmeans_obj} becomes a leading-order approximation to the variation of chemical source term using only the distance vector between two points in composition space as input (the primary input is are unchanged from standard K-means). If such an objective function can be minimized, the resulting clusters would identify regions of \textit{dynamical similarity} through the assignment matrix ${\bf L}^{\bf A}$, where dynamical similarity is defined as the degree to which two points in a local neighborhood can be described the same source term. This becomes clearer when observing that setting the scaling matrix as the identity (e.g. ${\bf A}_k = {\bf I}$ for all $k$, which amounts to stationary dynamics) restores the standard K-means algorithm.

Before proceeding, it should be noted that the general idea of modifying the standard K-means algorithm with weighting functions is not new. However, the JSK-means method introduced here differs from previous variants through reliance on a centroid-dependent scaling matrix and extraction of a scaling matrix using underlying physics-based dynamics. The reader is directed to \ref{app:kmeans_variants} for more detail on this matter.

\subsection{Jacobian-Scaled K-means Algorithm}
\label{sec:jacobian_algorithm}
The JSK-means algorithm to minimize the modified objective in Eq.~\ref{eq:jacobian_kmeans_obj} is provided in Algorithm~\ref{alg:jacobian_kmeans}. This is very similar to the standard K-means procedure in Algorithm~\ref{alg:standard_kmeans}. The major modification is in step (b), which consists of updating the scaling matrices via computing the chemical Jacobians at the current centroid locations as per Eq.~\ref{eq:chemical_jacobian}. These Jacobian matrices are required in step (c), because the labeling mechanism in JSK-means comes from evaluating norms of the sample-centroid distance vectors scaled by the Jacobian matrix ${\bf A}_k$ (see Eq.~\ref{eq:jacobian_indicator}). Although centroid initialization can be accomplished in the same way as with standard K-means (e.g. via the K-means++ algorithm \cite{kmeanspp}), in this work, a burn-in procedure that takes the converged centroids of the standard K-means algorithm is used for initialization of the JSK-means centroids. Through this approach, the converged centroids produced by the JSK-means algorithm can be interpreted as a modifier to the standard K-means centroids, which allows one to pinpoint the effect of the influence of Jacobian scaling in an interpretable way (as discussed in Sec.~\ref{sec:results}).

\begin{algorithm}
\caption{Jacobian-scaled K-means Clustering}
\label{alg:jacobian_kmeans}
\KwData{ \\ 
    (1) Set number of clusters $K$\\
    (2) Set convergence error tolerance $\varepsilon_{tol}$ \\
    (3) Initialize dataset $\Phi$ to be clustered \\ 
    (4) Initialize centroids: ${\bf c}_{new} \gets [c_1, \ldots, c_K]$ \\ 
    (5) Initialize convergence criterion: $\varepsilon_{c} \gets \text{Inf}$ \\
}
\vspace{0.5\baselineskip}
\KwResult{ \\ 
    (1) Converged set of centroids in ${\bf c}_{new}$. \\
    (2) Cluster assignment matrix ${\bf L}^{\bf A}$ (Eq.~\ref{eq:jacobian_indicator}) \\
}
\vspace{0.5\baselineskip}

\While{$\varepsilon_{c} \geq \varepsilon_{tol}$} 
{
    a) Copy centroids: ${\bf c}_{old} \gets {\bf c}_{new}$ \\ 
    b) Evaluate Jacobians (scaling matrices) at centroids in ${\bf c}_{old}$ (Eq.~\ref{eq:chemical_jacobian}): ${\cal A} \gets \{{\bf A}_1, \ldots, {\bf A}_K \}$ \\
    c) Update cluster assignment matrix ${\bf L}^{\bf A}$ using $\cal A$ and ${\bf c}_{old}$ (Eq.~\ref{eq:jacobian_indicator})\\
    d) Update centroids in ${\bf c}_{new}$ using cluster assignment matrix ${\bf L}^{\bf A}$ (Eq.~\ref{eq:jacobian_centroid}) \\ 
    e) Compute convergence criterion: $\varepsilon_c \gets \lVert {\bf c}_{new} - {\bf c}_{old} \rVert_F$  \\ 
    f) Break if specified number of iterations is reached
}
\end{algorithm}

Note that in Alg.~\ref{alg:jacobian_kmeans}, the definition of the centroid during the iterative convergence procedure has not changed from the standard K-means approach, but the assignment mechanism has changed by means of a different distance function. This produces an inconsistency, because the way in which the centroid update rule is derived is intrinsically tied to the distance function used in the definition of the K-means objective. Since the distance function has changed in the Jacobian-scaled approach, and the definition of the centroid has not changed, this inconsistency translates to eliminating the convergence guarantees of the standard K-means algorithm. 

More specifically, in the standard K-means approach, the goal is to derive a centroid update rule based on the cluster labels generated from the previous set of centroids that results in a reduction to the objective function in Eq.~\ref{eq:standard_kmeans_obj}. It can be shown, through either gradient descent or expectation maximization (EM) techniques, that such an update rule recovers Eq.~\ref{eq:standard_centroid} \cite{bengio_kmeans,murphy_book}. 

In JSK-means, however, the objective function is different from standard K-means. A derivation of the centroid update rule that minimizes Eq.~\ref{eq:jacobian_kmeans_obj} instead of Eq.~\ref{eq:standard_kmeans_obj} therefore yields a slightly different expression. If the updated centroid is denoted $c_{k,new}$, this update rule is 
\begin{equation}
    \label{eq:true_centroid_update}
    c_{k,new} = \frac {\sum_{i=1}^{N} {\bf L}^{{\bf A}}_{i,k} \phi_i } { \sum_{i=1}^{N} {\bf L}^{{\bf A}}_{i,k} } + R.
\end{equation}
In the above equation, the true centroid update for minimizing the JSK-means objective in Eq.~\ref{eq:jacobian_kmeans_obj} departs from the one used in Alg.~\ref{alg:jacobian_kmeans} by a residual $R$. Outlined in \ref{app:kmeans_derivation}, this residual is proportional to both the within cluster variance and the rate-of-change of the matrix ${\bf A}_k$ in the $D$-dimensional phase space (e.g., the chemical Hessian matrix). In this work, the residual in Eq.~\ref{eq:true_centroid_update} is ignored, allowing the JSK-means algorithm to retain the original centroid definition at the cost of guaranteeing a monotonic decrease of the objective function. As demonstrated in Sec.~\ref{sec:results}, good convergence trends in the JSK-means objective are still observed in practice via Alg.~\ref{alg:jacobian_kmeans}. Strategies for replacing the standard centroid update with the true update defined by Eq.~\ref{eq:true_centroid_update} will be explored in future work.

\subsection{Normalization Procedure for Jacobian Regularization}
\label{sec:scaling}
A normalization procedure is carried out before clustering to nondimensionalize the components of $\phi$ and the Jacobian matrices. This ensures that composition variables have equal contribution in both standard and Jacobian-scaled K-means algorithms. The normalized composition sample, denoted $\widetilde{\phi}_i$, is recovered from the raw $\phi_i$ via
\begin{equation}
    \label{eq:normalization}
    \widetilde{\phi}_i = {\bf B} \phi_i = 
    \begin{bmatrix} 
    \frac{1}{\max(\phi^1) - \min(\phi^1)} & &\\
    & \ddots & \\
    & & \frac{1}{\max(\phi^D) - \min(\phi^D)}
    \end{bmatrix}
    \begin{bmatrix} 
    \phi^1_i\\
    \vdots\\
    \phi^{D}_i
    \end{bmatrix}.
\end{equation}
Superscripts in Eq.~\ref{eq:normalization} denote component indices (not powers). The diagonal matrix ${\bf B}$ normalizes each respective component of $\phi$ by its range. An analogous scaling matrix, denoted $\bf C$, can be obtained to range-normalize chemical source term values. 

Given the diagonal scaling matrices for composition values and source terms (${\bf B}$ and ${\bf C}$ respectively), the scaled Jacobian matrix, denoted $\widetilde{\bf A}$, { can be recovered as}
\begin{equation}
    \label{eq:jacobian_scaling}
    \widetilde{\bf A} = {\bf D} \odot {\bf A}, \quad {\bf D} = \text{diag}({\bf C}) \text{diag}({\bf B}^{-1})^\text{T}, 
\end{equation}
where $\odot$ denotes an elementwise matrix product (Hadamard product) and the $\text{diag}(\cdot)$ operation produces a column vector containing the diagonals of the input matrix. The procedure in Eq.~\ref{eq:jacobian_scaling} ensures that the individual components in the Jacobian are scaled in a manner consistent with problem dimensionality without needing to find the ranges of individual Jacobian matrix elements.

It should be noted, however, that obtaining the Jacobian scaling matrix $\bf D$ does require estimates for the ranges (minimum and maximimum values) of both composition and chemical source terms, which in practice are extracted from an input dataset. The disadvantage here is that these scaling matrices will be application or problem dependent (this is true for any method that leverages a scaling approach in a preprocessing stage). Expert guided knowledge can be used to obtain ranges for composition values (e.g. temperatures in detonations typically fall within the range of 300K to 6000K), but it is important to acknowledge that the ranges for species and source term values cannot be prescribed this way. 

\begin{figure}
    \centering
    \includegraphics[width=0.5\columnwidth]{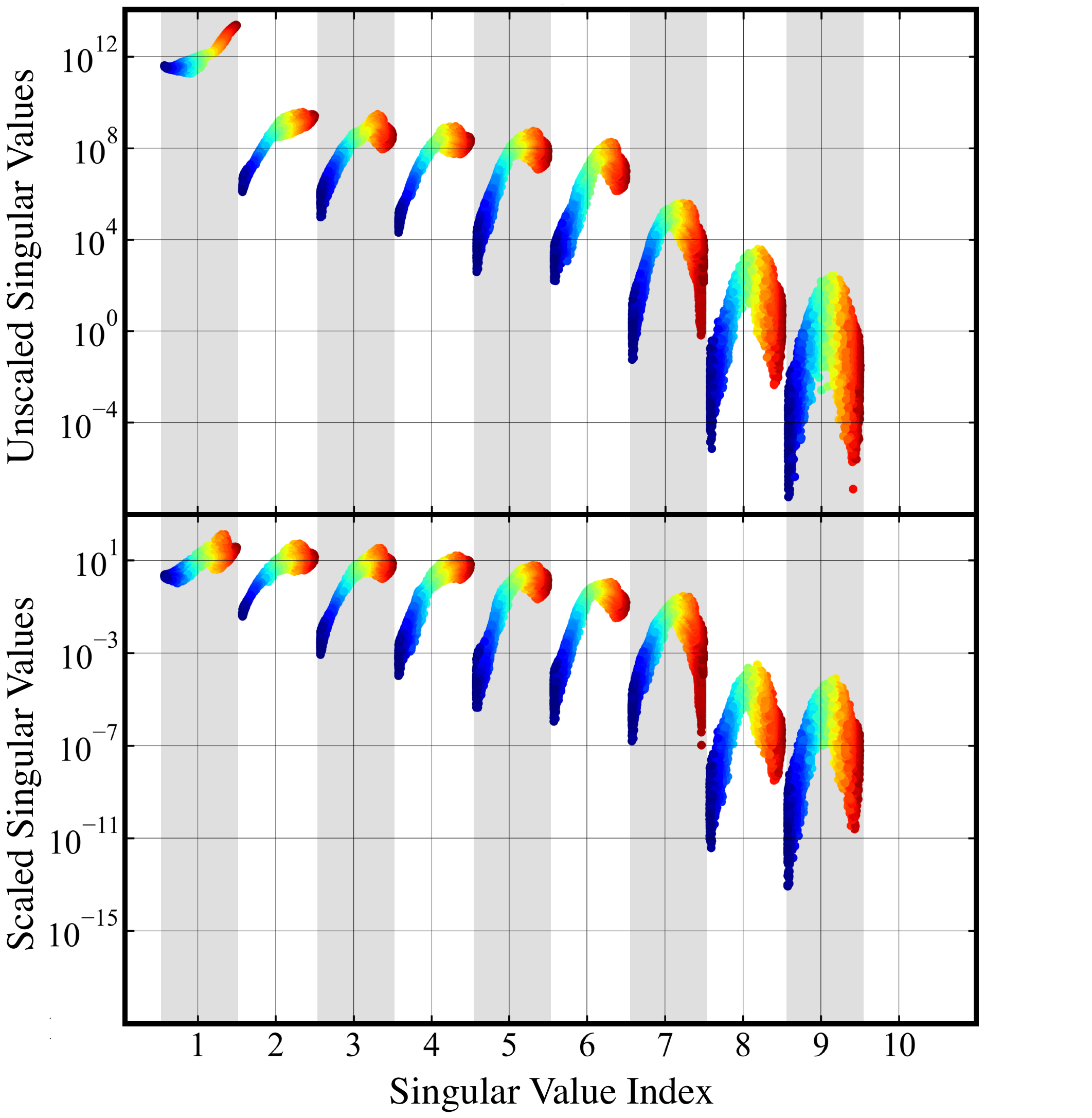}
    \caption{Singular value distribution for un-normalized (top) and normalized (bottom) chemical Jacobians obtained from dataset $\Phi$. The x-axis denotes the singular value index. For each index, spread in corresponding singular value for all $N$ sample points in $\Phi$ is plotted and colored by temperature. Points correspond to samples above 500 K. Bands are alternately shaded for ease of visualization.}
    \label{fig:singularvalues}
\end{figure}

The effect of the normalization procedure is evidenced in Fig.~\ref{fig:singularvalues}, which compares the singular value distributions of the scaled and unscaled chemical Jacobians. The figure shows how the scaling procedure drops the disparity between the largest and smallest (as well as the first and second) singular values in the data distribution to be used in the clustering procedure. The practical implication of this effect is a well-conditioned Jacobian matrix, leading to improved convergence of the JSK-means algorithm via Alg.~\ref{alg:jacobian_kmeans}. 

\section{Results} 
\label{sec:results}

\subsection{Demonstrative Toy Problem}
\label{sec:toy_problem}
Before proceeding with the detonation dataset, a demonstration of the JSK-means algorithm on a toy problem with a 1D feature space is first provided. { The intent is to describe how JSK-means, via the Jacobian-scaled distances used in Alg.~\ref{alg:jacobian_kmeans}, modifies the centroid locations and clusters produced by standard K-means on a simpler problem.} References to the general trends uncovered here are then made in Sec.~\ref{sec:detonation_results}, which performs extended analysis on the more complex reacting flow dataset. 

\begin{figure}
    \centering
    \includegraphics[width=\columnwidth]{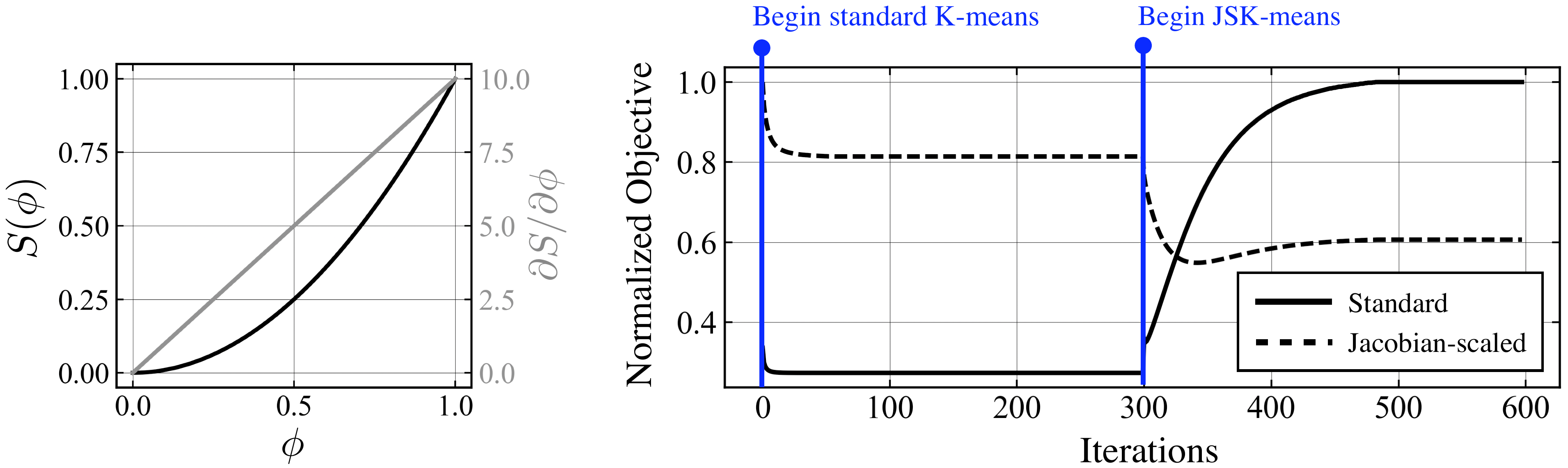}
    \caption{{ \textbf{(Left)} Plot of source term (black) and Jacobian (blue) versus phase space variable $\phi$ for the 1D toy problem in Eq.~\ref{eq:1d_toy_problem}. Variables in the plot have been normalized as per the procedure outlined in Sec.~\ref{sec:scaling}.} \textbf{(Right)} { Plot of normalized objective functions for standard (solid) and physics-guided (dashed) K-means clustering approaches versus number of iterations of the K-means algorithm.} The first 300 iterations is a burn-in phase that utilizes the standard K-means algorithm in Alg.~\ref{alg:standard_kmeans}. The next 300 iterations utilize the modified physics-guided K-means algorithm in Alg.~\ref{alg:jacobian_kmeans}.}
    \label{fig:1d_toy_problem}
\end{figure}

In this toy problem, the underlying nonlinear dynamical system is represented as a quadratic function via
\begin{equation}
    \label{eq:1d_toy_problem}
    \frac{d\phi}{dt} = S(\phi) = \frac{1}{2} \phi^2 \in \mathbb{R},
\end{equation}
which produces the trivial Jacobian $\partial S / \partial \phi = \phi$. A plot of the quadratic source term and linear Jacobian from Eq.~\ref{eq:1d_toy_problem} is shown in Fig.~\ref{fig:1d_toy_problem} (left). 

The input dataset is generated by sampling the phase space variable $\phi$ within the range $[0, 10]$ at $N=10000$ evenly spaced intervals. Figure~\ref{fig:1d_toy_problem} (right) plots the history of both standard and Jacobian-scaled K-means objective functions (Eqs.~\ref{eq:standard_kmeans_obj} and \ref{eq:jacobian_kmeans_obj}, respectively) using the burn-in procedure mentioned in Sec.~\ref{sec:jacobian_algorithm}, which consists of two stages. In the first stage (burn-in), the standard K-means algorithm is run for a prescribed number of iterations (300 here) using centroids initialized from the K-means++ procedure \cite{kmeanspp}; in the second stage, the JSK-means algorithm is run for the same number of iterations (or until the convergence criterion is met) using the converged centroids from standard K-means as the initial condition. In each of these phases, the objective function for both approaches is computed at each iteration. To facilitate comparison, the curves in Fig.~\ref{fig:1d_toy_problem} were then produced by normalizing each objective function by its observed maximum value during the entire iterative process. 
In the initial stage of the algorithm, which involves running the standard K-means algorithm for the first 300 iterations as depicted in Fig.~\ref{fig:1d_toy_problem} (right), a monotonic decrease in the standard K-means objective function is observed. This function (see Eq.~\ref{eq:standard_kmeans_obj}) measures the within-cluster sum of squares of $\phi$, and its steady decrease during the first stage is expected. Interestingly, the JSK-means objective function also decreases in a similar monotonic behavior in this stage, despite the fact that the standard K-means objective is being minimized. 

Iteration 300 indicates the onset of the JSK-means algorithm. At this point, the converged centroids from standard K-means are supplied as the initial centroids to JSK-means. After the switch, a noticeable drop in the JSK-means objective function is observed, while the standard K-means objective function rises. This trend is in-line with the goal of the JSK-means algorithm, as the JSK-means objective function approximates the within-cluster variation in $S(\phi)$ as opposed to $\phi$. This effect is indicative of nonlinearity in the source term, and captures a key feature of the JSK-means method: optimizing the cluster partitions for source term variation necessitates a tradeoff in the optimization of the clusters for variations in the state variable. Additionally, as evidenced in Fig.~\ref{fig:1d_toy_problem} and referenced in Sec.~\ref{sec:jacobian_algorithm}, the JSK-means algorithm does not guarantee monotonic decrease of the corresponding objective function (convergence occurs when the centroids are stabilized in the $D$-dimensional phase space). 

\begin{figure}
    \centering
    \includegraphics[width=0.8\columnwidth]{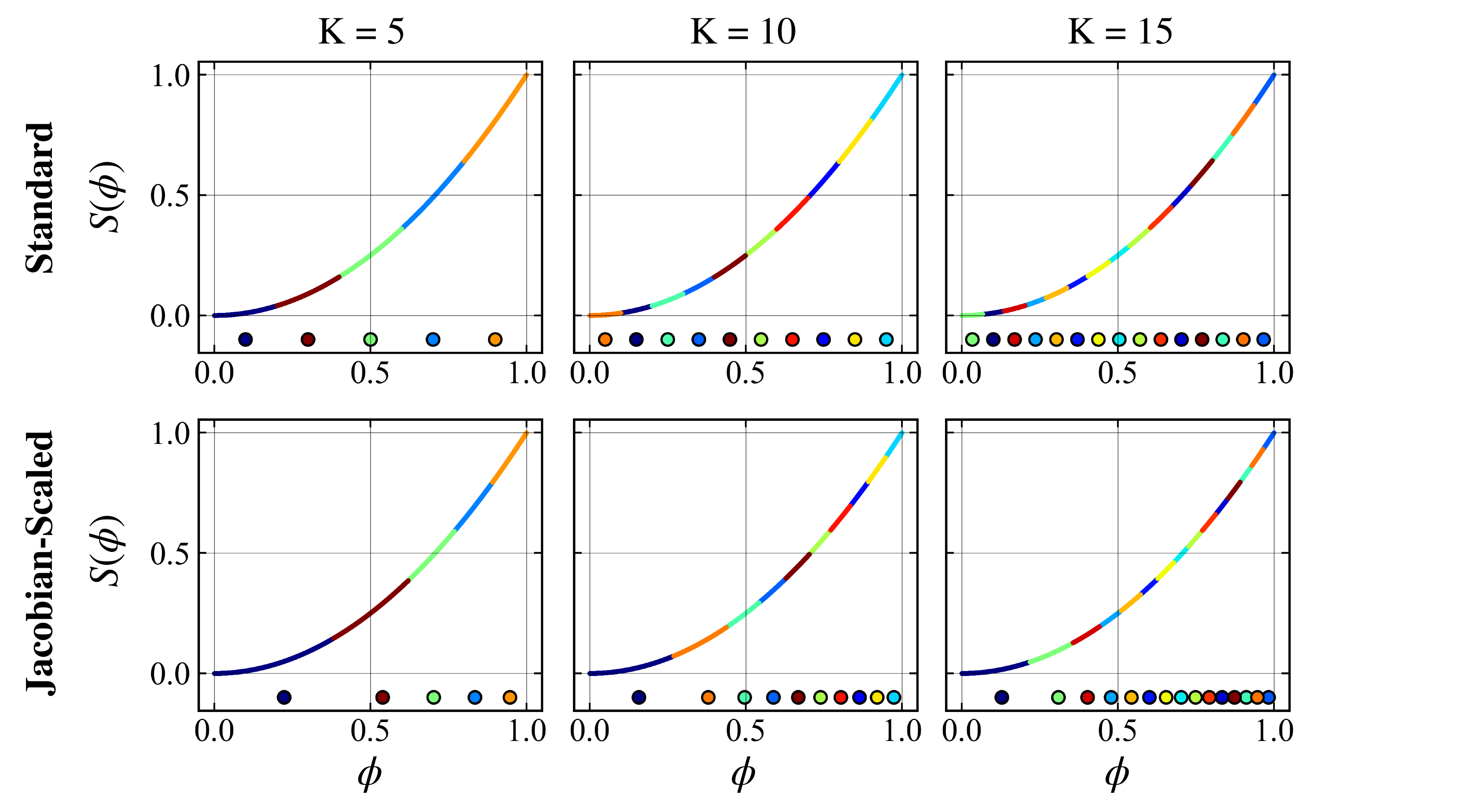}
    \caption{Cluster visualizations provided by standard K-means (top row) and JSK-means algorithm (bottom row) for $K=5$, $10$, and $15$. Centroid locations are provided as the filled markers in each plot.}
    \label{fig:1d_cluster_visualizations}
\end{figure}

\begin{figure}
    \centering
    \includegraphics[width=0.7\columnwidth]{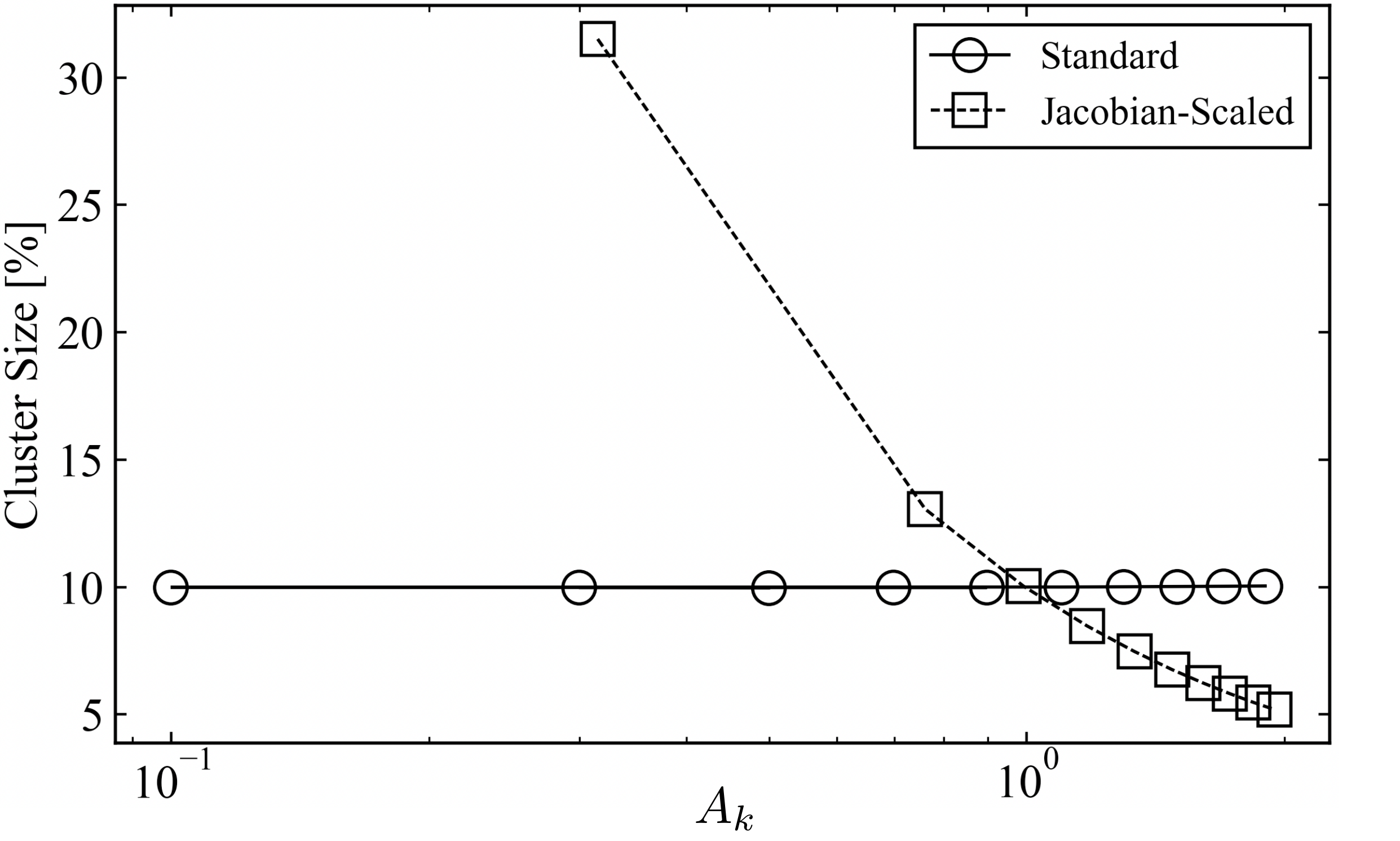}
    \caption{{Cluster size versus centroid Jacobian for standard K-means and JSK-means clusters in the 1D toy problem ($K=10$). Cluster size is defined as percentage of within-cluster samples to total number of samples.}}
    \label{fig:cluster_size}
\end{figure}

Ultimately, Fig.~\ref{fig:1d_toy_problem} shows how JSK-means drops the corresponding Jacobian-based objective function in Eq.~\ref{eq:jacobian_kmeans_obj} as intended, producing a set of converged centroids that reduce variation in nonlinear source terms within the clusters. To better interpret this effect, a visualization of the change in centroid locations provided by the JSK-means algorithm with reference to the converged centroids from the standard K-means algorithm is provided in Fig.~\ref{fig:1d_cluster_visualizations} for $K=5$, $10$, and $15$ clusters. { In all cases, Fig.~\ref{fig:1d_cluster_visualizations} shows how the JSK-means algorithm shifts and redistributes the standard K-means clusters towards high-sensitivity regions in the phase space.} This effect can be quantitatively accessed by correlating the cluster sizes with centroid Jacobian norms, as shown in Fig.~\ref{fig:cluster_size} -- the weighting towards high-sensitivity regions implies that clusters are redistributed in regions where there is significant nonlinearity. Put another way, for a finite set of $K$ centroids, the physics-guided clustering algorithm through Jacobian-scaling ensures that the $K$ centroids are localized in regions of dynamical similarity as prescribed by the governing equations via Eq.~\ref{eq:1d_toy_problem}. 

\subsection{Detonation Flowfield}
\label{sec:detonation_results}
The JSK-means clustering algorithm is now extended to the detonation dataset described in Sec.~\ref{sec:dataset}. More specifically, in a similar manner as in the toy problem above, convergence trends for various $K$ values will be analyzed and changes in cluster assignments in both physical space and composition space due to Jacobian-scaling will be interpreted in Sec.~\ref{sec:detonation_physical_space} and \ref{sec:detonation_comp_space}, respectively.

\subsubsection{Analysis of Clusters in Physical Space via Segmented Fields}
\label{sec:detonation_physical_space}

\begin{figure}
    \centering
    \includegraphics[width=0.6\columnwidth]{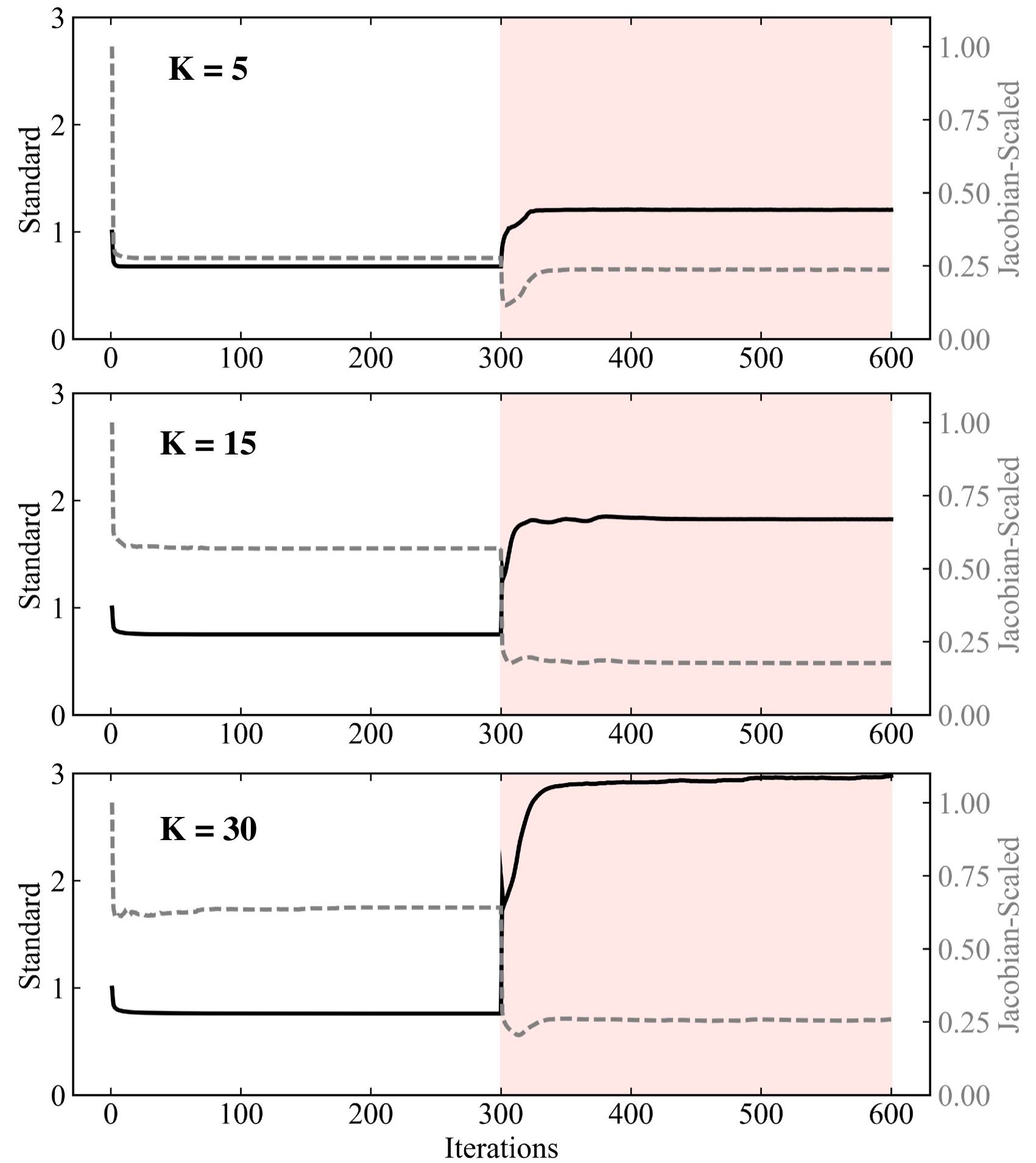}
    \caption{{ Standard K-means (black, Eq.~\ref{eq:standard_kmeans_obj}) and  JSK-means (gray, Eq.~\ref{eq:jacobian_kmeans_obj}) objective function values during the iterative procedure for $K=5$ (top), $K=15$ (middle), and $K=30$ (bottom). Curves are averaged over 15 different centroid initializations. The first 300 iterations is the burn-in period, where standard K-means is run for 300 iterations -- the red shaded region denotes the switch to the JSK-means algorithm in Alg.~\ref{alg:jacobian_kmeans}. \textit{To facilitate comparison at various $K$ values, objective function curves are relative with respect to the initial conditions; i.e., they are normalized by the corresponding values at the first iteration.}}}
    \label{fig:detonation_convergence}
\end{figure}

Before visualization and interpretation of the partitioned composition space, it is important to ensure that the algorithm presented in Alg.~\ref{alg:jacobian_kmeans} produces the a similar decrease in the JSK-means objective in the detonation dataset as {observed in the 1D toy problem} in Sec.~\ref{sec:toy_problem}. To this end, convergence trends for both standard (Eq.~\ref{eq:standard_kmeans_obj}) and Jacobian-scaled (Eq.~\ref{eq:jacobian_kmeans_obj}) objective functions are provided in Fig.~\ref{fig:detonation_convergence} for $K=5$, $15$ and $30$ (objective function curves are normalized in the figure to facilitate comparison across different cluster numbers). { For each cluster number, the curves represent an average from 10 different sets of initial centroids provided to the standard K-means algorithm to initiate the burn-in procedure, resulting in unique K-means runs.} Overall, the convergence trends in Fig.~\ref{fig:detonation_convergence} show how Alg.~\ref{alg:jacobian_kmeans} minimizes the physics-based Jacobian-scaled objective in Eq.~\ref{eq:jacobian_kmeans_obj} even for the complex detonation dataset. {As with the 1D problem}, there is no noticeable oscillatory behavior in the objective curves, indicating that the centroid locations have stabilized. 

\begin{figure}
    \centering
    \includegraphics[width=0.5\columnwidth]{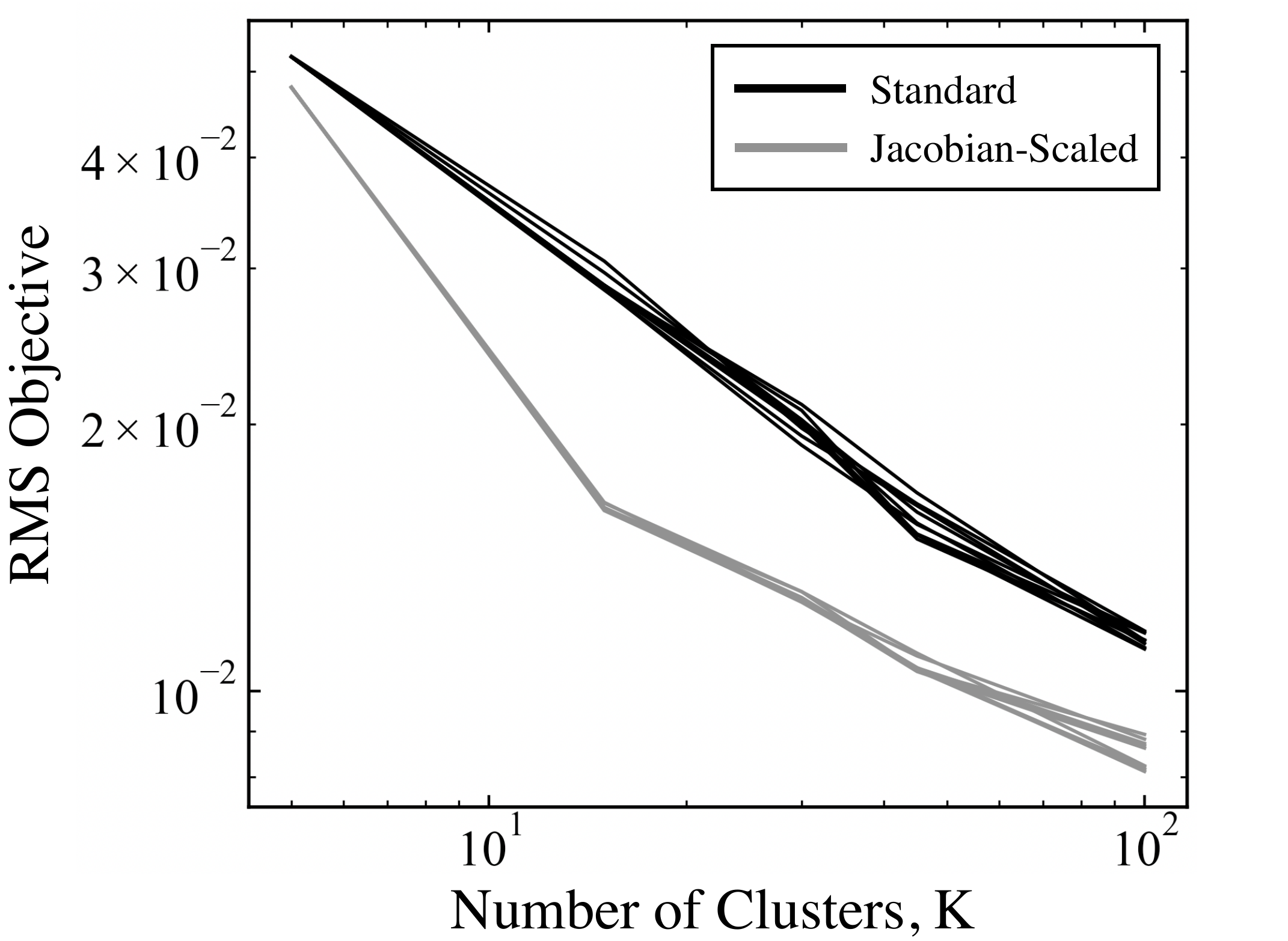}
    \caption{{ Evaluation of physics-based RMS objective function (Eq.~\ref{eq:rms_objective}) versus number of clusters using standard K-means output (black) and JSK-means output (gray).}}
    \label{fig:detonation_cluster_number}
\end{figure}

{Relative to the initial state, Fig.~\ref{fig:detonation_convergence} shows how the drop-off in the JSK-means objective increases noticeably from $K = 5$ to $15$; however, there is no noticeable modification of the degree of drop-off in the normalized objectives when jumping from $K = 15$ to $30$. The implication is that the reduction provided in the relative JSK-means objective stabilizes, but the reduction in the absolute values of the objective decreases, as described below. Interestingly, the upward jump in at iteration 300 in the normalized standard K-means objectives in Fig.~\ref{fig:detonation_cluster_number} (black curves) increases with $K$. This implies that at higher $K$, JSK-means more heavily modifies the standard K-means clusters -- relative to the initial condition -- to achieve the reduction to its objective function.}

Figure~\ref{fig:detonation_cluster_number} shows the effect of cluster number on the objective function in Eq.~\ref{eq:jacobian_kmeans_obj} computed from both converged standard K-means clusters and JSK-means clusters. For consistent comparison across cluster numbers, a normalized version of the Jacobian-scaled objective, termed the root-mean-square (RMS) objective, is plotted instead of the value in Eq.~\ref{eq:jacobian_kmeans_obj}. This RMS objective is given by 
\begin{equation}
    \label{eq:rms_objective} 
    E_{{\bf A}}^{RMS} = \sqrt{ \frac{1}{N} \sum_{k=1}^{K} \sum_{i=1}^{N} 
    \big \lVert 
    {\bf L}^{{\bf A}}_{i,k}  {\bf A}_k  [\phi_i - c_k]
    \big \rVert_2^2 }.
\end{equation}
Recall that the initial set of centroids provided as input to the JSK-means algorithm comes from the converged standard K-means centroids. As such, to produce the curves in Fig.~\ref{fig:detonation_cluster_number}, the above RMS objective is computed using centroid and distance evaluations from both standard and Jacobian-scaled approaches. The trends in Fig.~\ref{fig:detonation_cluster_number} show how the modification of centroid locations provided by the JSK-means approach results in decreased within-cluster variations of source term. An increasing cluster number in both standard and JSK-means approaches results in a nearly constant-rate reduction in the Jacobian-based RMS objective, which is indicative of the localization of clusters. Interestingly, despite the fact that the standard K-means approach does not optimize Eq.~\ref{eq:rms_objective}, higher cluster numbers still result in lower values of Eq.~\ref{eq:rms_objective}, implying that source terms are localized in composition space for this dataset. Usefully, even at higher cluster numbers (e.g. $K=100$), the gap in RMS objective between the JSK-means and standard K-means evaluations remains, which signifies that the JSK-means algorithm succeeds in shifting the baseline clusters produced by standard K-means towards regions of greater dynamical similarity. 

\begin{figure}
    \centering
    \includegraphics[width=0.7\columnwidth]{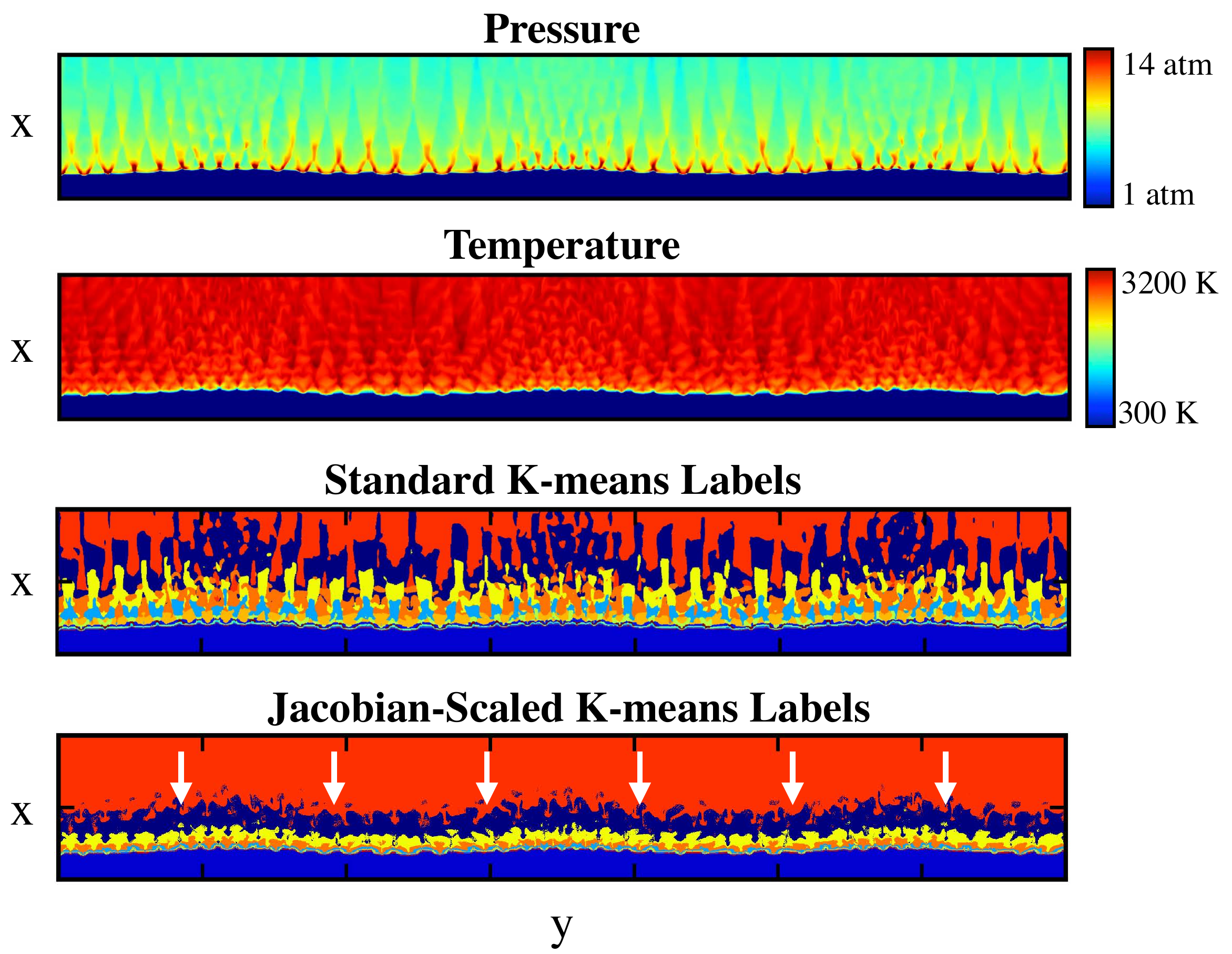}
    \caption{From top-to-bottom: pressure, temperature, standard K-means labels, and JSK-means labels for detonation dataset as described in Sec.~\ref{sec:dataset}. Flowfields have been transposed (wave is moving towards bottom of page) for ease of visualization.}
    \label{fig:segmented_field_full}
\end{figure}

Although informative, analysis of trends in the objective function values must be supplemented with a visual inspection of the cluster partitions in physical space for a clearer interpretation of the impact of the JSK-means clustering procedure. To this end, cluster assignments obtained from standard K-means and JSK-means are shown in Fig.~\ref{fig:segmented_field_full} for the $K=15$ case -- general trends discussed hereafter apply to all studied $K$ values. Note that the actual colors of the cluster labels in physical space come from the cluster index, and are not meaningful from a physical perspective -- however, coherent regions in physical space defined by cells of the same color (or cluster) are indicators for key flow features extracted by the unsupervised algorithms. The cluster label plots in physical space are termed the standard (obtained from standard K-means) and Jacobian-scaled (obtained from JSK-means) segmented fields respectively.

The segmented fields in Fig.~\ref{fig:segmented_field_full} reveal the way in which the JSK-means algorithm biases, or pushes, the cluster labels towards the wavefront and detonation reaction zone (indicated by the white arrows in the respective plot). This is consistent with the fact that the difference between the standard and Jacobian-scaled approach is in the chemical Jacobian based scaling of the distance function, which necessitates the localization (or redistribution) of centroids towards regions of high composition sensitivity. This phenomenon is directly analogous to the biasing and weighting of the centroids seen in the toy problem in Sec.~\ref{sec:toy_problem}. { As a result, the redistribution of centroids in composition space provided by JSK-means translates to localizing the cluster partitions near regions of high chemical heat release, which can be a useful property for developing more accurate models that intend to target such complex regions in the simulation procedure (see Sec.~\ref{sec:applications_and_need}).} On the other hand, the standard K-means approach produces a segmented field that is significantly more refined and complex in turbulence-dominated regions far behind the wavefront that are unimportant from the chemical contribution perspective.

\begin{figure}
    \centering
    \includegraphics[width=0.8\columnwidth]{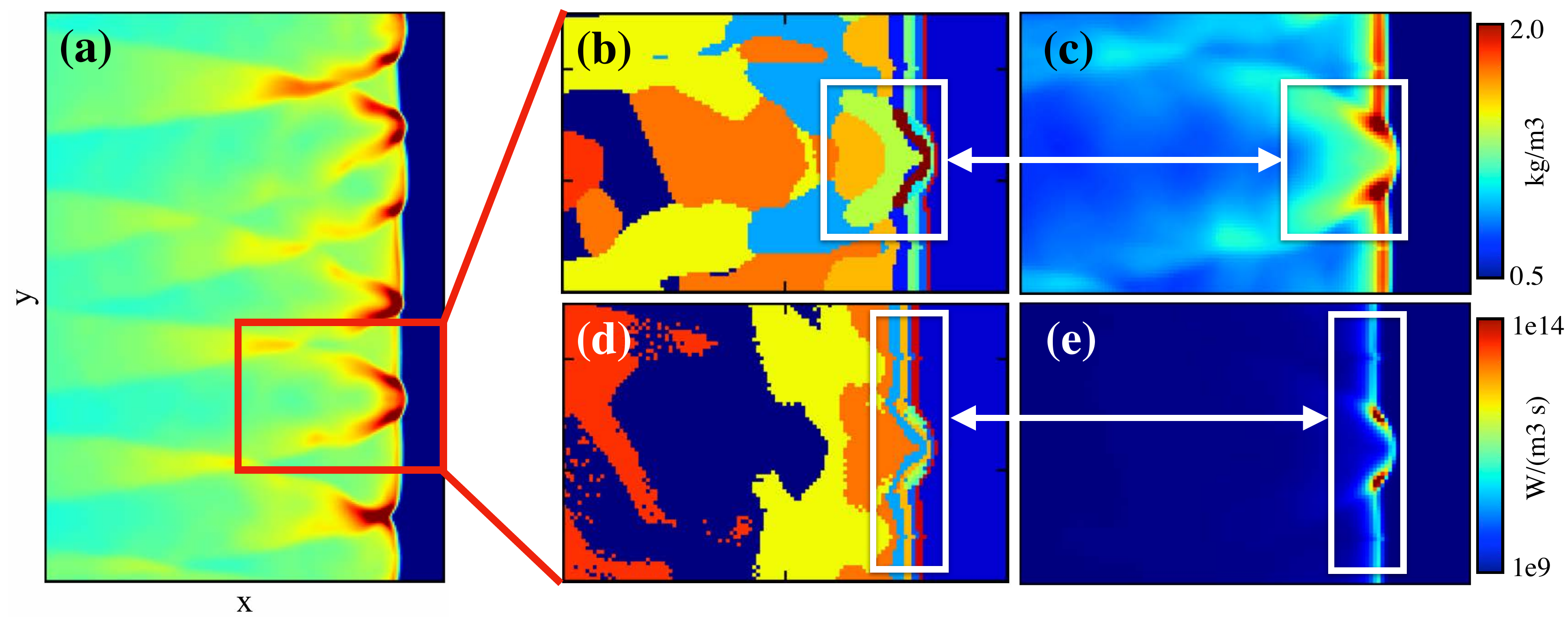}
    \caption{\textbf{(a)} Pressure profile within detonation wave in the domain window of $x=[0.087, 0.094]$~m and $y=[0.04, 0.05]$~m (coordinate axes supplied in Fig.~\ref{fig:snapshot}). Red box indicates zoom-in region on triple point structures for remainder of plots in the figure. \textbf{(b)} Standard K-means labels in triple point region.  \textbf{(c)} JSK-means labels in triple point region. \textbf{(d)} Fluid density (kg/$\text{m}^3$) in the triple point region. \textbf{(e)} Heat release rate (W/$\text{m}^3$/s) in the triple point region. White boxes in (b)-(e) indicate correspondence in respective segmented field structure and key flow features.}
    \label{fig:segmented_field_zoom}
\end{figure}

For a better visualization of the algorithm behavior near the wavefront, Fig.~\ref{fig:segmented_field_zoom} shows a series of zoomed-in profiles of pressure, density, heat release rate, and the same segmented fields from Fig.~\ref{fig:segmented_field_full}. Before assessing the near-wavefront trends in the segmented fields, it should be noted that the pressure profile shown in Fig.~\ref{fig:segmented_field_zoom} displays the characteristic triple-point structures that oscillate vertically along the wavefront as it propagates through the channel. In short, triple points are localized high-pressure regions at the detonation wave front emanating from collisions between a weaker series of reflecting transverse waves and the leading shock wave which travels at the CJ speed \cite{shepherd_proci}. The triple point structures, indicated by the small distributed regions of peak pressure throughout the wavefront, contribute to significant complexity in the detonation wave dynamics and are known to heavily influence chemical kinetic behavior within the detonation wave structure \cite{venkat_arfm_rde}. As such, part of assessing the strength of any composition space partitioning algorithm for detonation-containing flows is to ensure that areas near and within the triple point structures are detected by the given algorithm with ideally minimal user input. 

The differences between the segmented fields in this triple point region are quite apparent in the zoom-ins -- for example, in the standard K-means segmented field, the cluster partitions attempt to recover spatially coherent patterns that align more with density fields and contact discontinuities away from the wavefront, { which contribute to the complex and less spatially-uniform cluster partitions observed downstream.} This comes from the definition of the composition vector, which consists of temperature as well as density-weighted mass fractions (i.e. species concentration); as such, there is an intrinsically higher weight placed on non-reacting transverse wave propagation in the standard clustering procedure, which in turn leads to complex 2-dimensional structures recovered in the segmented field.

{On the other hand, the segmented fields produced by the JSK-means approach are markedly more 1D in overall structure} (e.g. the variation in cluster transitions along the y-direction is noticeably smaller than that observed in the standard K-means segmented fields). The effect of biasing clusters towards the wave front, which was visualized macroscopically in Fig.~\ref{fig:segmented_field_full}, is also apparent in the zoom-ins. The JSK-means algorithm pushes clusters towards the regions of high chemical sensitivity, as encoded in corresponding regions of peak heat release rate and the triple points. Because the contact discontinuities and transverse wave oscillations behind the detonation wavefront produce much less chemical reactivity as implied by the heat release rates, the cluster partitions from the Jacobian-scaled approach are more vertically uniform behind the wavefront. { Additionally, there is a higher degree of noise in the Jacobian-scaled segmented field than in the standard counterpart (i.e., there is a high-frequency scatter effect slightly away from the wavefront, indicated by the red regions in in Fig.~\ref{fig:segmented_field_zoom}(d)) -- this may be due to potentially high variations in chemical stiffness, as well as possible numerical errors attributed to numerical chemical Jacobian estimates for intermediary species.}

\subsubsection{Analysis of Clusters in Composition Space via Proper Orthogonal Decomposition}
\label{sec:detonation_comp_space}

The above discussion is geared towards interpretation of the output of the JSK-means clustering procedure in relation to the standard K-means approach in physical space. However, additional insight into the workings of JSK-means can be extracted by visualizing the cluster distributions in composition space. Because the composition space is high-dimensional ($D=10$ here), direct visual analysis of cluster partitions is not straightforward. For interpretable composition space analysis, projections onto two-dimensional spaces facilitated by proper orthogonal decomposition (POD) will be utilized. It is assumed that the reader is already familiar with the basics of POD (and equivalently, principal component analysis) -- fundamentals and background on the approach are provided in Refs.~\cite{pca_1987,berkooz_pod} and the references therein.

To better visualize the effects of JSK-means, two sets of POD bases are derived: one from the dataset containing the composition variables $\Phi$, and another from the dataset containing the source terms $\Omega$. { It should be noted that before carrying out the POD procedure, each of these datasets here were range-scaled as per the procedure described in Sec.~\ref{sec:scaling}.}

\begin{figure}
    \centering
    \includegraphics[width=0.5\columnwidth]{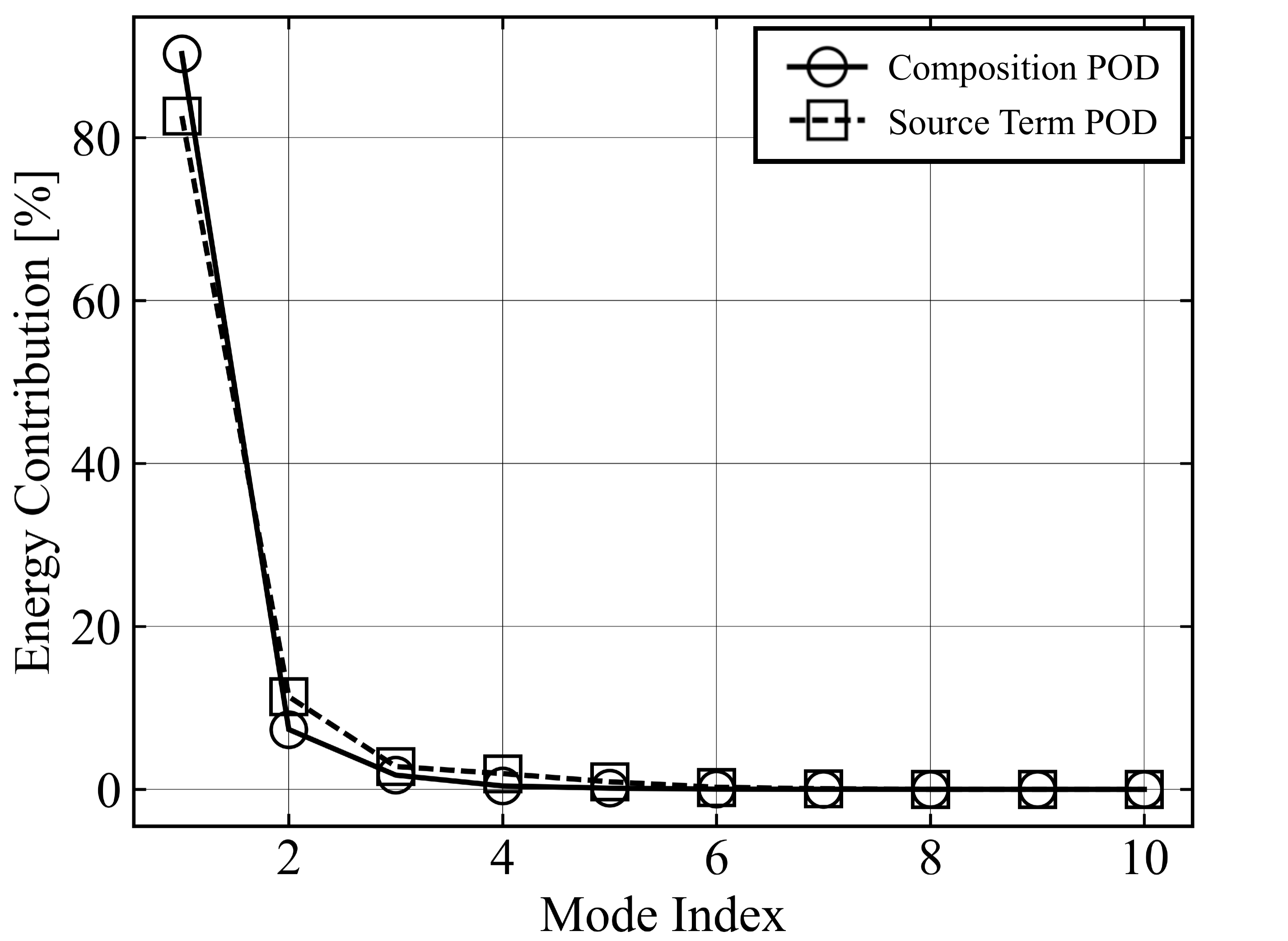}
    \caption{Energy contribution versus mode index, measured as a percentage of total variance captured by the POD modes, from the phase/composition space decomposition (Eq.~\ref{eq:phase_coefficients}) and source term decomposition (Eq.~\ref{eq:source_coefficients}).}
    \label{fig:pod_energy_distribution}
\end{figure}

The POD basis derived from $\Phi$ is denoted ${\bf U}_{\Phi} \in \mathbb{R}^{D \times M}$, and the POD basis derived from  $\Omega$ is denoted ${\bf U}_{\Omega} \in \mathbb{R}^{D \times M}$. The columns of the matrices ${\bf U}_{\Phi}$ and ${\bf U}_{\Omega}$ contain the respective basis vectors (also known as principal component directions), and $M$ denotes the number of retained modes in the expansion. Projection of the datasets onto their respective POD basis vectors produces the POD coefficients, the components of which are uncorrelated due to the orthonormal property of the basis. More formally, the phase space coefficients are obtained from the projection
\begin{equation}
    \label{eq:phase_coefficients}
    {\bf A} = [a_1, \ldots, a_N] = {\bf U}_{\Phi}^{\text{T}} {\Phi}  \in \mathbb{R}^{M \times N},
\end{equation}
and the source term coefficients from the analogous projection
\begin{equation}
    \label{eq:source_coefficients}
    {\bf B} = [b_1, \ldots, b_N] = {\bf U}_{\Omega}^{\text{T}} {\Omega} \in \mathbb{R}^{M \times N}. 
\end{equation}
In the above equations, the $a_i$ and $b_i$ are $M$-dimensional column vectors for the $i$-th phase space and source term coefficient, respectively. For two samples $i$ and $j$, if all basis vectors are retained in the POD representation, the distance between two points $\phi_i$ and $\phi_j$ in composition space can be cast as
\begin{equation}
    \lVert \phi_i - \phi_j \rVert_2^2 = \lVert a_i - a_j \rVert_2^2, 
\end{equation}
and similarly, the distance between the same two points in a so-called "source-term" space can be cast as 
\begin{equation}
    \lVert S(\phi_i) - S(\phi_j) \rVert_2^2 = \lVert b_i - b_j \rVert_2^2.
\end{equation}
The above property comes from the fact that the data variance (and Euclidean distance) is preserved in the POD expansion if all modes are retained \cite{berkooz_pod}. Mode "energies" extracted from the eigenvalues of respective covariance matrices can then be used to quantify the percentage contribution of each POD mode to the overall data variance. This is shown in Fig.~\ref{fig:pod_energy_distribution}; usefully, for both composition and source term data projections, the first two POD modes retain most of the data variance. As such, 2-dimensional visualizations utilizing only the first two POD coefficients from both the composition dataset (Eq.~\ref{eq:phase_coefficients}) and the source term dataset (Eq.~\ref{eq:source_coefficients}) can be reliably used to directly assess the effects of the JSK-means clustering procedure in terms of centroid distributions and cluster labels. 

\begin{figure}
    \centering
    \includegraphics[width=\columnwidth]{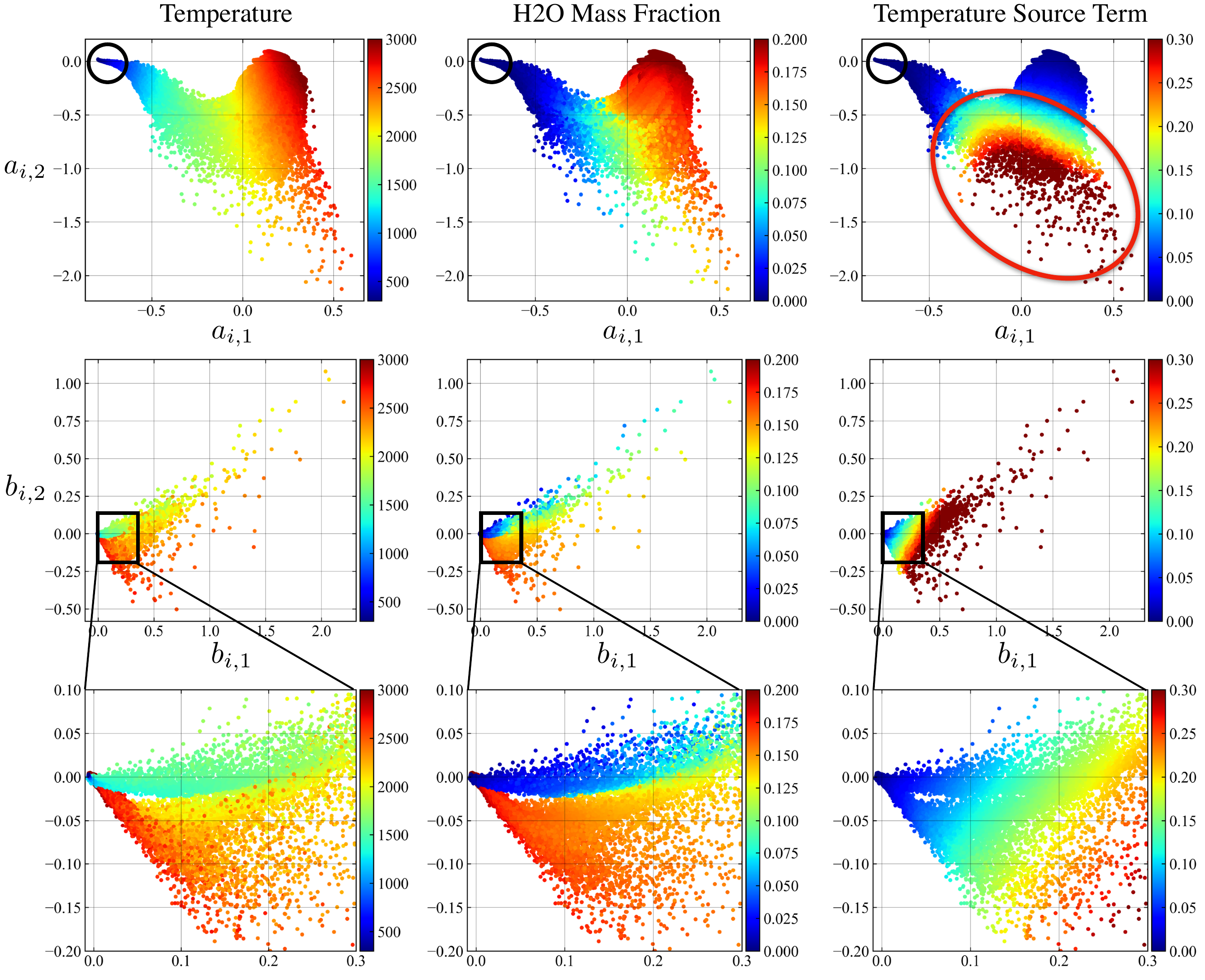}
    \caption{{ \textbf{(Top row)} From left-to-right, visualization of temperature field in units of Kelvin, H2O mass fraction, and temperature source term (nondimensionalized as per Sec.~\ref{sec:scaling}) in the two-dimensional composition POD coordinates (Eq.~\ref{eq:phase_coefficients}). Black circle indicates ambient region (unreacted gas ahead of the detonation wave), and red circle indicates regions of high chemical heat release at reactivity (high-sensitivity regions within the detonation wave structure). \textbf{(Middle row)} Same as top row, but for source term POD coordinates (Eq.~\ref{eq:source_coefficients}). \textbf{(Bottom row)} Zoom-ins of source term POD coordinates.}}
    \label{fig:pod_scatter}
\end{figure}

Figure~\ref{fig:pod_scatter} shows examples of these visualizations in both the composition space coefficients ($a_i$, Eq.~\ref{eq:phase_coefficients}) and source term coefficients ($b_i$, Eq.~\ref{eq:source_coefficients}). There is significant correlation between the POD coefficients and key quantities of interest, such as temperature fields, H$_2$O mass fraction (which can be interpreted as a reaction progress variable), as well as chemical source terms; these correlations confirm that the respective POD projections facilitate a concise representation of data variance in both composition and source term space in two dimensions. Note that in the source term space (bottom row of Fig.~\ref{fig:pod_scatter}), the regions in composition space (top row) that produce zero chemical source term collapse to a single point (the $(0,0)$ coordinate). Usefully, the correlation in temperature source term becomes much more linear in the source term space as opposed to the composition space -- in other words, an increase in values in the first principle component of the source term POD coefficient produces, to reasonable confidence, an increase in chemical reactivity. 

\begin{figure}
    \centering
    \includegraphics[width=\columnwidth]{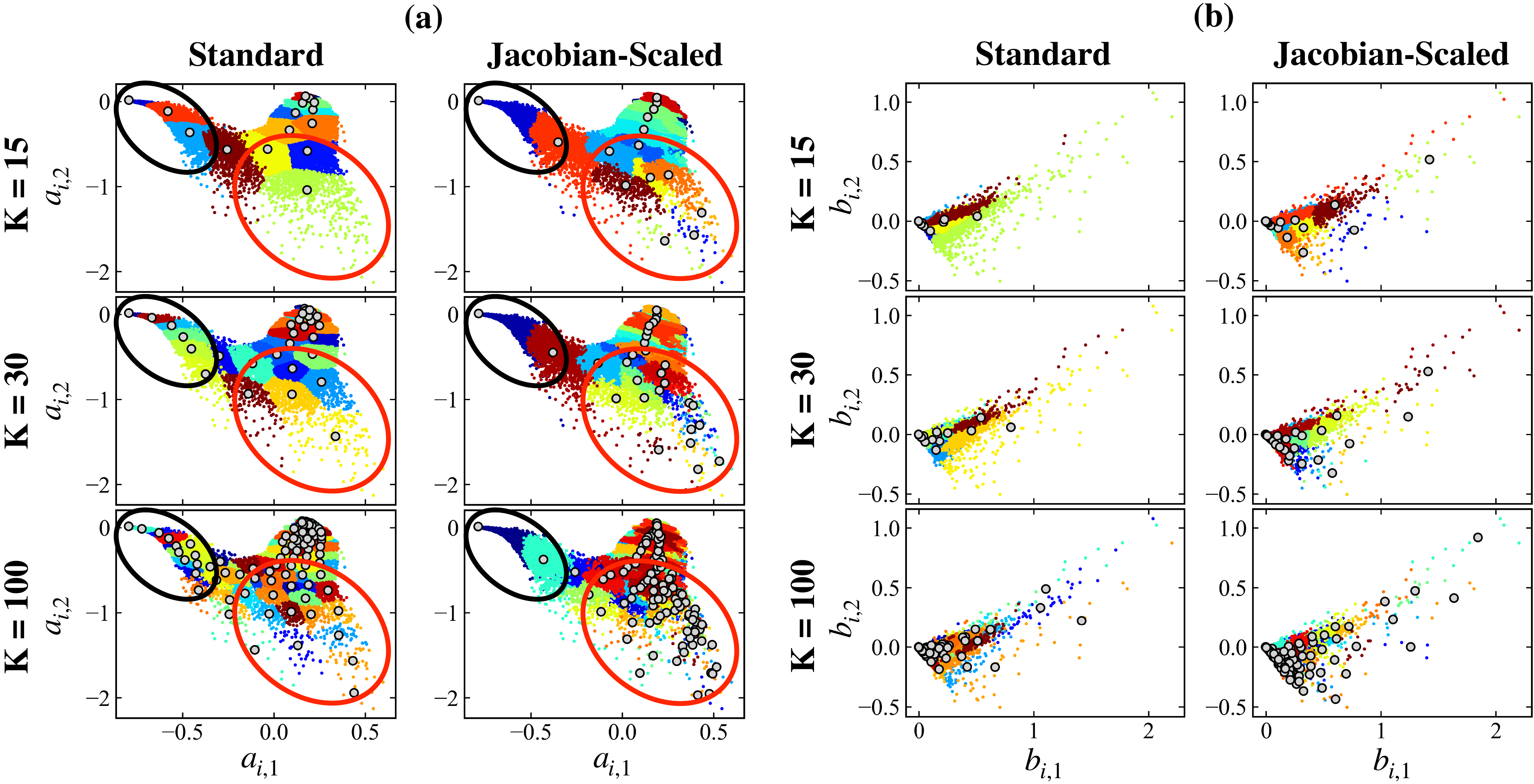}
    \caption{\textbf{(a)} Visualizations of cluster labels from standard K-means (left column) and JSK-means (right column) in the two-dimensional composition POD space (Eq.~\ref{eq:phase_coefficients}) for $K=15$, $30$, and $100$. Data samples are colored based on cluster index, and centroids are denoted by gray markers outlined in black. Large black circle outlines chemically non-reacting and ambient regions, whereas red circle outlines chemically reacting/sensitive regions at or near the detonation wave front. \textbf{(b)} Same as (a), but centroids and cluster labels are plotted in the source term POD space (Eq.~\ref{eq:source_coefficients}).}
    \label{fig:pod_labels}
\end{figure}

Figure~\ref{fig:pod_labels}(a) compares the centroid locations and corresponding cluster labels produced by the standard K-means with those produced by JSK-means in the two-dimensional composition POD space. Figure~\ref{fig:pod_labels}(b) displays the same quantities in the source term POD space. The plots illustrate how JSK-means refines clusters in fundamentally different regions of phase space than in the standard approach. More specifically, as indicated in Fig.~\ref{fig:pod_labels}(a), the JSK-means algorithm successfully redistributes centroids away from the ambient, non-reacting regions (black circles) and towards the highly reactive regions (red circles) that characterize much of the complexity in the detonation wave structure (i.e. regions of peak heat release rates). { With regards to cluster refinement, the physics-guided approach of JSK-means refines clusters in these same reactive regions, which is equivalent to allocating additional clusters near the detonation wavefront and triple point structures where high chemical heat release and species sensitivity is expected to occur.}

In the dual sense, increasing the number of clusters in the Jacobian-scaled approach does not result in a refinement of the ambient, non-reacting region of the flowfield, which is not true for standard K-means. This effect of centroid redistribution is particularly apparent in the source term POD space projections of Fig.~\ref{fig:pod_labels}(b), which shows how JSK-means increases the amount of centroid spread in source term space; this in turn drops the within-cluster variation of source term distances within each cluster. The trends in Fig.~\ref{fig:pod_labels} both support the segmented field behavior of the JSK-means clustering approach in physical space (Figs.~\ref{fig:segmented_field_full} and \ref{fig:segmented_field_zoom}), and also allude to the fact that the modified algorithm in Alg.~\ref{alg:jacobian_kmeans} is indeed dropping the modified objective in Eq.~\ref{eq:jacobian_kmeans_obj}. 

{ \section{Applications and Need for JSK-Means} 
\label{sec:applications_and_need}
The discussion above showcased the impact of JSK-means in the clustering procedure by means of biasing (or altering) centroid positions in feature space to reflect dynamically sensitive regions. In the context of chemically reacting flows, with knowledge of the chemical source term, application of JSK-means to composition space clustering is shown in the above sections to effectively steer centroids towards regions of high source term sensitivities. The goal of this section is to (a) provide a brief overview of some practical use-cases of this approach (Sec.~\ref{sec:use_cases}), and (b) provide justification as to why JSK-means is useful from the modeling perspective, instead of using standard K-means clustering on the source terms directly (Sec.~\ref{sec:source_term_clustering}). 

\subsection{JSK-means Use Cases}
\label{sec:use_cases}
The JSK-means approach has two primary use-cases (although there can be more): (1) facilitating physical analysis of complex flow fields (e.g., turbulent reacting flows), and (2) partition-based (or tabulation-based) surrogate modeling. By construction, these use-cases are inherited directly from those of standard K-means and other partition-based clustering approaches. The key advantage here is that the same partition-based modeling and analysis approaches can be constructed using JSK-means clusters as conditioning variables, which necessarily take into account the composition space dynamics. Both aspects are discussed in the following. 

    \underline{Cluster-based physical analysis:} K-means, when interpreted as a data decomposition tool, has been used in a wide variety of contexts to drive physical analysis of fluid flows (the reader is directed to Refs.~\cite{kaiser_crom,shivam_timeaxis} and the references therein for different takes on this). Although the scope of this work is primarily concerned with introducing JSK-means as a way to inject a physics-based inductive bias into standard K-means, the utility of JSK-means from the physical analysis perspective, through construction of segmented flowfields, should not be overlooked. In other words, the act of assigning a sample (which here represents a composition vector as well as a physical space location) to a cluster ID, and subsequently visualizing the resulting cluster IDs in physical space through the segmented fields, is itself a useful application of the approach, especially when comparing with standard K-means outputs. Such a comparison offers a highly interpretable avenue for breaking down complex reacting flowfields in a physically meaningful way; unlike standard K-means, the JSK-means segmented fields reveal dynamically similar regions from the perspective of the chemical source term, and allow the user to directly connect these regions to coherent structures in physical space. In unsteady turbulent reacting flow contexts, such visualizations (i.e., the time evolution of the segmented flowfield) can play a meaningful role in investigating flow physics from high-fidelity datasets, particularly from the angle of flow-chemistry interactions. 

    To demonstrate this point, such analysis of the JSK-means clusters can give insight into regions described by fast/prohibitive chemical timescales. For example, Fig.~\ref{fig:timescales} shows average chemical timescales -- both maximum and minimum -- in each cluster produced by standard K-means and JSK-means at K=15 and 30 for the channel detonation configuration. Interestingly, through inspection of these average timescales, it is evident that the Jacobian-based weighting of JSK-means pushes clusters towards smaller chemical timescales (i.e., many of the K-means clusters with minimum chemical timescales above $10^{-8} s$, with the exception of outliers representing the ambient region, are eliminated). The inset in Fig.~\ref{fig:timescales} shows that, for the smallest values of minimum timescales, a much larger range of maximum timescales is also captured by JSK-means (clusters not only tend towards smaller timescales, but encompass a larger range of stiffness). This shift in the timescale distribution of JSK-means clusters coincides with the segmented field concentration around the wavefront and triple points in the detonation structure described in Sec.~\ref{sec:detonation_results}, which in turn can facilitate further physical analysis and feature extraction. To further emphasize this point and as an additional demonstration of the utility of JSK-means, the reader is directed to Ref.~\cite{michael_jskmeans}, where JSK-means is used to analyze high-speed combustor flowfields in a full-scale configuration. Here, despite a fundamentally different configuration, chemical timescale effects are consistent with those observed in Fig.~\ref{fig:timescales}. 

    \begin{figure}
        \centering
        \includegraphics[width=\columnwidth]{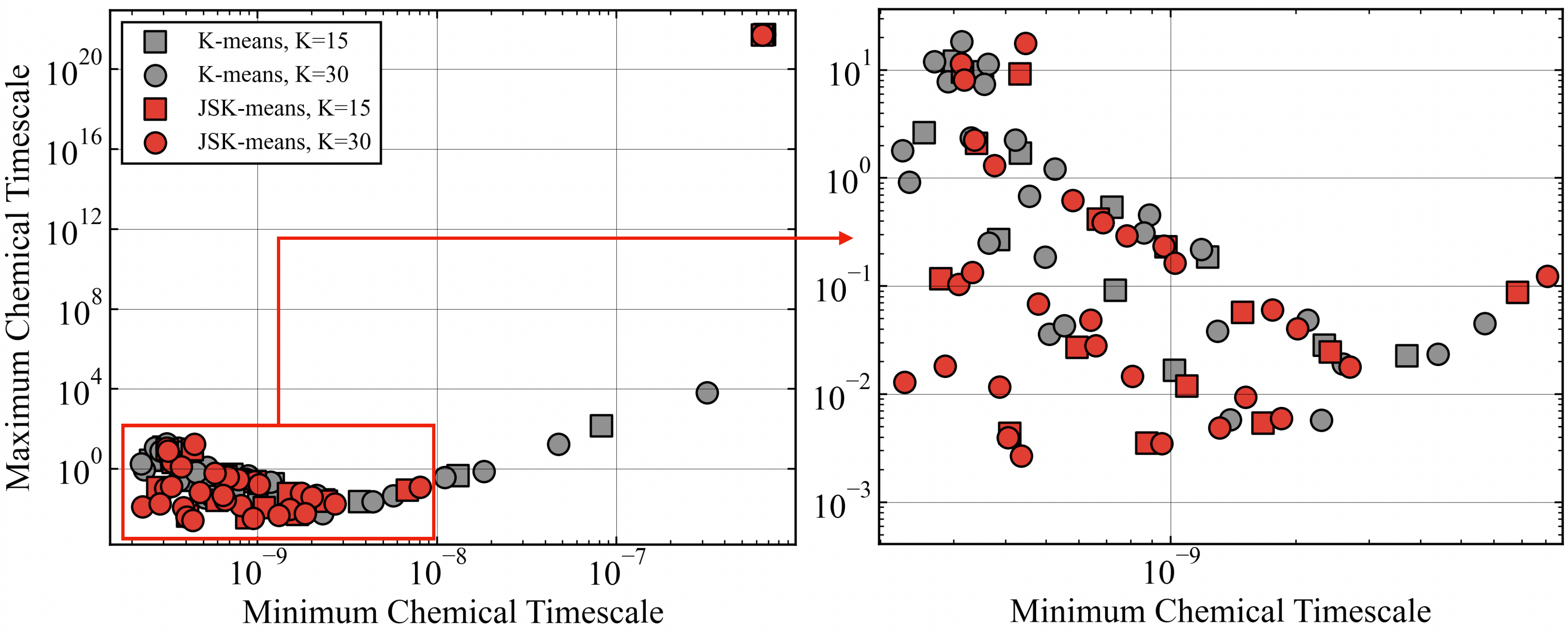}
        \caption{Maximum versus minimum chemical timescales in each cluster (timescales in units of seconds) for K=15 (squares) and K=30 (circles). Gray points denote standard K-means, red points denote JSK-means. Markers reflect averages of cluster samples.}
        \label{fig:timescales}
    \end{figure}

    \underline{Surrogate modeling:} A primary motivation for the JSK-means approach is to move towards improved partition-based models that serve as accelerated surrogates for unknown or known but high-cost function evaluations. An example of such a function is the chemical source term $S(\phi)$. Partition-based modeling approaches condition the model evaluation on a localized representation of the state space, which amounts to the cluster ID of the query sample in cluster-based modeling workflows (examples of works that leverage partition-based modeling are provided in the Introduction). Since such workflows are typically tasked with modeling dynamics, investigation of model evaluations conditioned on JSK-means partitions, which are informed of the source terms themselves, is warranted. For demonstrative purposes, a simple JSK-means based modeling strategy for the chemical source term is provided and compared to a standard K-means counterpart below. 

    Given a general non-linear function $\psi = F(\phi)$, it is assumed that the objective of the modeling is to develop an (ideally accelerated) surrogate model for $F$, denoted $\widetilde{F}$, where $\widetilde{\psi} = \widetilde{F}(\phi)$. A \textit{cluster-based} surrogate then requires conditioning the model $\widetilde{F}$ on the cluster ID of a given input sample $\phi$ -- i.e., $\widetilde{\psi} = \widetilde{F}(\phi; k)$, where $k$ is the cluster (coming from either K-means or JSK-means) in which the input $\phi$ resides. To demonstrate the advantages of JSK-means, the demonstration here focuses on the combustion modeling task, where $F$ is equal to the instantaneous chemical source term $S$, such that $\psi=d\phi/dt$, and $\widetilde{S}= \widetilde{F}$ is a model for the source term. However, it should be emphasized that $F$ can represent other functions, such as the action of a stiff time integrator \cite{pope_isat} -- in this case, $\psi$ represents the composition at a future time instead of the instantaneous rates. 

    As a simple demonstration, a cluster-conditioned model is cast as the Taylor expansion about the $k$-th centroid (see Eq. 11) via
    \begin{equation}
        \widetilde{S}(\phi; k) =  S(c_k) + {\bf A}_k (\phi - c_k),
    \end{equation}
    where ${\bf A}_k$ is the Jacobian evaluated at the corresponding centroid, $S(c_k)$ is the source term evaluated at the centroid, and $(\phi - c_k)$ is the sample-to-centroid distance. This resembles a tabulation-based approach, since ${\bf A}_k$ and $S(c_k)$ are stored at the centroid locations and thereby provide the necessary cluster conditioning. For the detonation dataset, source term predictions using both standard K-means and JSK-means clusters are provided in Fig.~\ref{fig:linear_model} for a subset of the source term variables using $K=15$ and $K=30$ partitions. Due to the simplicity of this linear model, it must be acknowledged that overall prediction accuracy using both clustering strategies { leaves room for improvement}. However, the figure shows how JSK-means predictions provide noticeable improvement over standard K-means. Interestingly, this performance improvement is larger at lower cluster numbers; prediction accuracy gains are especially apparent at higher values of the temperature source term and throughout all minor species source term ranges. 
    
    These demonstrations ultimately point to the utility of JSK-means from the modeling standpoint: a clustering strategy informed of underlying dynamics naturally has potential in improving on partition-based models in which the partitions are not produced with knowledge of the dynamics (i.e., standard K-means and other similar clustering approaches). It is emphasized, however, that the $\widetilde{S}(\phi; k)$ formulation above is a simple linear model, and further investigation of JSK-means models using more expressive formulations -- such as neural networks \cite{shivam_ftc}, for example -- would complete the picture.

    \begin{figure}[ht]
        \centering
        \includegraphics[width=\columnwidth]{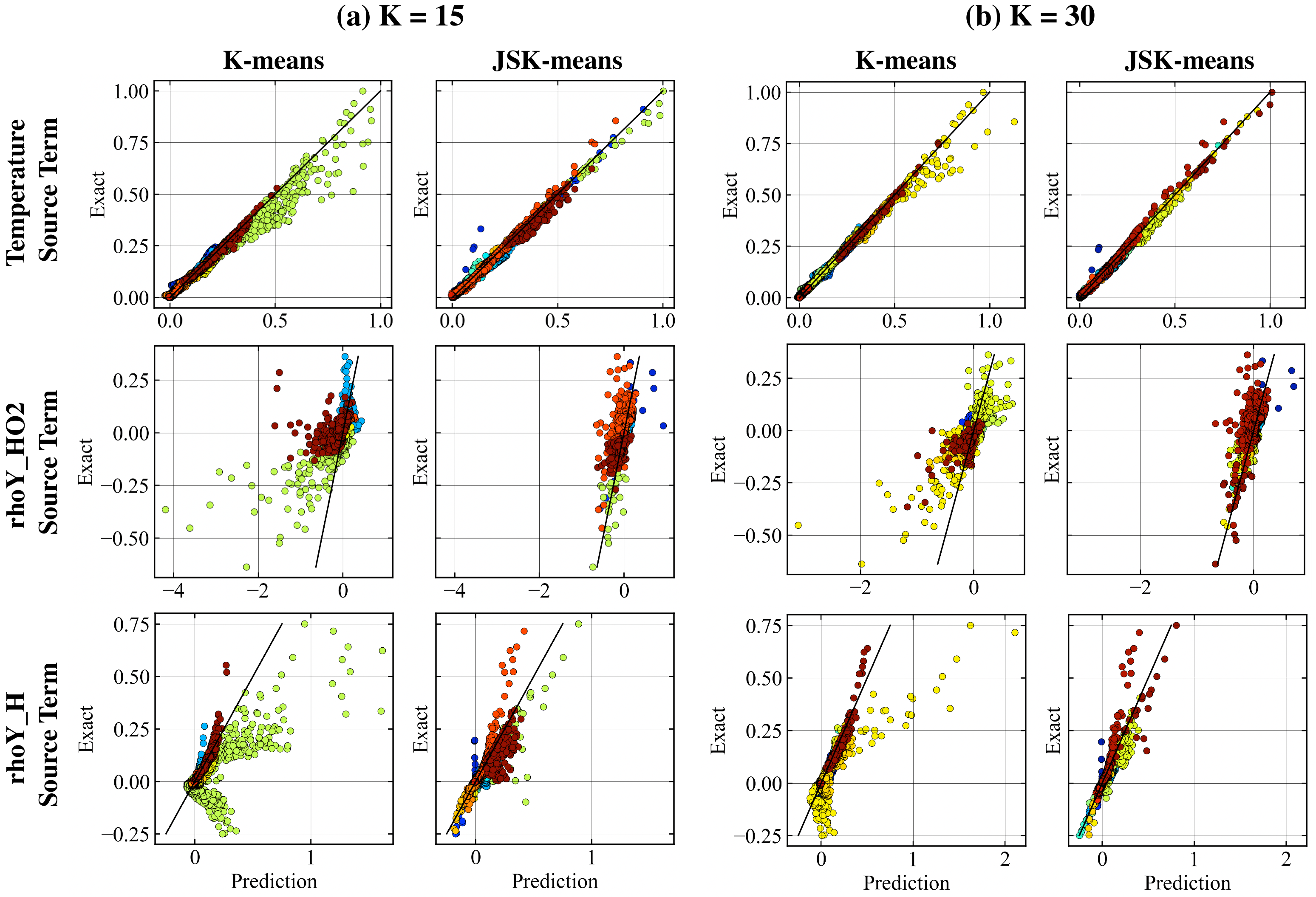}
        \caption{\textbf{(a)} K-means versus JSK-means based source term predictions using linear expansion model for K=15. Rows show scaled temperature, $\rho Y_{\text{HO}2}$, and $\rho Y_{\text{H}}$ source terms. In each plot, x-axis is predicted value ($\widetilde{S}(\phi)$), y-axis is target value (${S}(\phi)$), and markers of the same color indicate same cluster ID. \textbf{(b)} Same as (a), but for K=30.}
        \label{fig:linear_model}
    \end{figure}

    \subsection{Distinction from Source Term Clustering}
    \label{sec:source_term_clustering}
    The methodology of JSK-means begs the following question: \textit{What is the advantage of modifying the objective function using the Jacobian over executing standard K-means on the chemical source terms directly?} 

    Indeed, executing a clustering over the source terms directly ensures that resulting clusters are dynamically similar by design -- usage of such clustering in a physical analysis setting is an interesting direction. However, the JSK-means approach offers advantages over straight source term clustering, particularly when considering downstream modeling/integrated inference settings. These are described below.
    
    JSK-means strikes a balance between localization of clusters in composition space (standard K-means on $\phi$) and biasing towards regions of high reaction sensitivity (standard K-means on $S(\phi)$). This is a by-product of the linear Taylor expansion: the Jacobian-scaled distances ${\bf A}_k (\phi - c_k)$ recover the source term distance $(S(\phi) - S(c_k))$ as the sample $\phi$ becomes infinitesimally close to the centroid $c_k$. { The main takeaway and primary advantage in JSK-means is that the algorithm operates on the \textit{same} input data as standard K-means and appends knowledge of the state space dynamics to ultimately shift (or bias) the cluster distributions towards dynamics-oriented regions}. As a result, JSK-means centroids still represent localized averages of the input samples, as in standard K-means, but the clusters themselves are biased as per the Jacobian scaling -- as such, clusters are still localized in composition space. { This is not true when running standard K-means on the source terms $S(\phi)$, which is fundamentally different -- this produces centroids that represent conditionally averaged source terms.} 
    
    Considering the above implications, since JSK-means centroids represent conditional averages in composition space, JSK-means allows for the creation of tabulation-based localized combustion models (described in Sec.~\ref{sec:use_cases}). In other words, a potential use-case of the algorithm is a scenario in which all of the source term data for $S(\phi)$ (which again can represent either instantaneous source term evaluations or chemical time integration) is unavailable and needs to be modeled. In such a setting, clustering over $S(\phi)$ directly is not an option, as this is the modeling target. 
}

\section{Conclusion}
In this work, a physics-informed variant of K-means clustering, termed Jacobian-scaled K-means (JSK-means) clustering, was introduced. The method allows for the injection of underlying physical knowledge into the clustering procedure through a distance function modification: instead of leveraging conventional Euclidean distance vectors, the JSK-means procedure operates on distance vectors scaled by Jacobian matrices. These Jacobian matrices are cluster-dependent and are derived from dynamical system Jacobians (right-hand-side sensitivities) in the phase space of samples. Jacobian-scaled distances enter the clustering procedure through modified definitions of the clustering objective and cluster assignment functions. To arrive at the JSK-means algorithm, updates to the baseline K-means approach based on these modifications are made, including (a) computing Jacobians at the current centroid locations during the iterative procedure, and then (b) using these updated Jacobians to scale respective sample-to-centroid distance vectors. The general goal of this work is to show how, with this simple modification, the JSK-means algorithm -- without modifying the input dataset -- produces clusters that capture regions of dynamical similarity, in that the clusters are redistributed towards high-sensitivity regions in phase space, and are described by similarity in the source terms of samples instead of the samples themselves. 

The JSK-means algorithm was demonstrated in this work on a complex reacting flow simulation dataset. In this application, each sample to be clustered is described by the local thermochemical composition (species concentrations and temperature) extracted from a computational cell in a flowfield containing a self-sustained hydrogen-air detonation wave. The dynamics in this composition space are known through the highly nonlinear and stiff Arrhenius-based chemical source terms. As such, the distance scaling required by JSK-means was accomplished using cluster-dependent chemical Jacobian matrices evaluated at the cluster centroids.

With this dataset, detailed comparisons between outputs from JSK-means and standard K-means were made to highlight the practical effects of including Jacobian-scaled distances in the clustering procedure. To this end, convergence of the JSK-means algorithm based on an initial condition prescribed by the output of a standard K-means procedure (the burn-in approach) resulted in a redistribution, or biasing, of centroids towards regions of high chemical sensitivity. More specifically, it was shown how the modified algorithm successfully drops the corresponding JSK-means objective function during the iterative procedure, at the cost of increasing the standard K-means objective function as a tradeoff. Visualization of this effect in physical space was performed via the analysis of segmented fields -- { the flowfield delineations produced by the JSK-means algorithm pushed standard K-means clusters towards regions of high peak heat release rate near the detonation reaction zone}. Further visualization of this effect in composition space via POD projections showed how the Jacobian scaling separates (or maximizes variation between) clusters in the source-term space, without explicitly using source term data during the clustering procedure.

The objective of this work was to both introduce the JSK-means algorithm and show how it can be successfully used to generate dynamically consistent partitions in complex reacting flow applications. The results shown here demonstrate the promising capability of the JSK-means approach to improve upon previous partition-based models for chemical source terms (and other quantities of interest) that were made using clustering approaches that did not have knowledge of the governing equations (e.g., Ref.~\cite{shivam_ftc}). This type of modeling extension is actively being explored by the authors and will be reported in future work. { Additional avenues for future work include detailed investigations on the effect of the number of clusters $K$ on JSK-means convergence properties, coupling Jacobian-scaled distances with other clustering algorithms outside of the K-means framework, { comparing JSK-means clustering with direct source term clustering using standard K-means,} and evaluating the method in other multi-physics applications.}

\section*{Declaration of Competing Interests}
The authors declare that they have no known competing financial interests or personal relationships that could have appeared to influence the work reported in this paper.

\section*{Data Availability}
Data and source code will be made available on request.

\section*{Acknowledgements} 
The authors acknowledge support through ONR Grant No. N00014-21-1-2475 with Dr. Eric Marineau as Program Manager. S.B. acknowledges support by the Argonne Leadership Computing Facility, which is a U.S. Department of Energy Office of Science User Facility operated under contract DE-AC02-06CH11357.

\bibliographystyle{elsarticle-num} 
\bibliography{references}

\appendix

\section{Differences between JSK-means and Other K-means Variants}
\label{app:kmeans_variants}
The general idea of modifying the standard K-means algorithm with weighting functions is not new. In the class of density-based K-means clustering strategies, instead of modifying the distance function directly, the standard phase space distance in Eq.~\ref{eq:standard_distance} is scaled with an input PDF $\rho(\phi_i)$ \cite{kmeans_densityweighted}, which effectively biases the centroid locations to regions where this PDF is high. In other approaches, the distance vector $\delta \phi = \phi_i - c_k$ is scaled by a global diagonal matrix $\bf D$, producing a modified distance vector $\widetilde{\delta \phi} = {\bf D} \delta \phi$ that prioritizes certain components (or features) of $\phi$ over others in the distance evaluation, where the diagonal elements of $\bf D$ are estimated during the clustering optimization procedure \cite{huang_weighted_kmeans,tseng_penalized_kmeans}. This amounts to a scaling operation on the phase space vector, and these approaches are typically constrained to fixing the same diagonal matrix $\bf D$ for all clusters, the advantage being that doing so leads to no required changes in the convergence properties of the baseline K-means clustering algorithm. Additionally, the class of kernel K-means strategies reformulates the distance computation by transforming the phase space variable $\phi \in \mathbb{R}^D$ into another variable $\psi \in \mathbb{R}^B$ \cite{kernel_kmeans,deep_kmeans}. Distances are evaluated in the $\psi$-space by means of either the kernel trick or latent-space transformations. 

A comparison of the above K-means variations with the JSK-means approach presented here is shown in Table.~\ref{tab:kmeans_objective_comparison}. The primary differences are (a) the scaling matrix ${\bf A}_k$ is a function of the cluster index, which means that during an iterative optimization procedure based on the convergence of centroid locations, the ${\bf A}_k$ must be updated as the centroids change, (b) the scaling matrices are derived from underlying physical rules via the right-hand-side Jacobian of the dynamical system, which ensures that the clusters adapt to highly sensitive regions in the phase space, and (c) the definition of the centroid as the arithmetic mean of within-cluster samples (Eq.~\ref{eq:jacobian_centroid}) is the same as the baseline algorithm.

\begin{table}
\scriptsize
\centering
\begin{tabularx}{\columnwidth} 
  {
  | >{\centering\arraybackslash}X 
  | >{\centering\arraybackslash}X 
  | >{\centering\arraybackslash}X 
  | >{\centering\arraybackslash}X 
  | >{\centering\arraybackslash}X | 
  }
 \hline
 \textbf{K-means Variant} & \textbf{Sample-Centroid Distance} & \textbf{Description} & \textbf{Key References} \\
 \hline \hline 
 Standard K-means  & $\lVert \phi_i - c_k  \rVert_2^2$ & Standard Euclidean distance & \cite{kmeans_main_reference} \\
 \hline
 Jacobian-scaled K-means & $\lVert {\bf A}_k [\phi_i - c_k] \rVert_2^2$ & Cluster-dependent scaling matrix derived from dynamical system Jacobian & This work \\
 \hline
 Diagonal Weighted K-means & $\lVert {\bf D} [\phi_i - c_k] \rVert_2^2$  & Global diagonal scaling matrix & \cite{huang_weighted_kmeans} \\
 \hline
 Density-based K-means & $\rho(\phi_i) \lVert \phi_i - c_k \rVert_2^2$  & Scaling of standard Euclidean distance via PDF evaluation & \cite{kmeans_densityweighted} \\
 \hline
 Kernel / Deep K-means  & $\lVert \psi_i - c_k \rVert_2^2$  & Standard Euclidean distance evaluated in latent space  & \cite{kernel_kmeans,deep_kmeans} \\
 \hline
\end{tabularx}

\caption{Comparison of K-means variations with the Jacobian-scaled K-means approach.}
\label{tab:kmeans_objective_comparison}
\end{table}

\section{Derivation of Centroid Update Rule}
\label{app:kmeans_derivation}

The sections below outline derivations of the centroid update rule for three versions of the K-means objective: 

\begin{enumerate}
    \item The standard K-means objective that utilizes the vanilla Euclidean distance measure (Sec.~\ref{sec:app:standard_kmeans}).
    \item A modified K-means objective that scales sample-centroid distances by a fixed scalar value that is independent of the centroid locations (Sec.~\ref{sec:app:constant_scaled_kmeans}).
    \item A modified K-means objective that scales sample-centroid distances by a centroid-dependent scaling variable (Sec.~\ref{sec:app:jacobian_kmeans}); this is the same modification used in the Jacobian-scaled K-means approach in Sec.~\ref{sec:jskmeans}. 
\end{enumerate}
As will be seen below, the derivations show that the centroid update rule obtained from minimizing (1) and (2) comes from the average of within-cluster samples -- i.e. scaling the K-means objective by a constant factor does not change the standard centroid update rule as expected. However, for approach (3), which is the same modification used in the Jacobian-Scaled K-means formulation in Sec.~\ref{sec:jacobian_algorithm}, minimization of the modified objective produces a centroid update rule that deviates from the within-cluster average by a residual proportional to the within-cluster variance of samples.

For illustrative purposes and to simplify the derivations, the input samples here are scalars (i.e. $D=1$); derivation trends and main takeaways from the scalar case are expected to apply in general to higher dimensional cases. Also, to facilitate clearer and more readable derivations, the notation used below deviates from that used in Sec.~\ref{sec:methodology}, and instead follows the notation used in Ref.~\cite{bengio_kmeans}. 

The input data samples are denoted $\phi_i \in \mathbb{R}$, $i=1,\ldots,N$, where $N$ is the number of samples. The centroids are denoted $c_k \in \mathbb{R}$, $k=1,\ldots,K$, where $K$ is the number of clusters. The set of centroids is given by $c = \{c_1, c_2, \ldots, c_K \}$.

The objective in the K-Means algorithm is to find a $c$ that minimizes the within-cluster variance. This can be expressed in the obective function $E(c)$, where 
\begin{equation}
    \label{eq:app:standard_objective}
    E(c) = \sum_i \frac{1}{2} \norm{ \phi_i - c_{s_i(c)} }^2_2 = \sum_i \frac{1}{2} \left( \phi_i - c_{s_i(c)}\right)^2.  
\end{equation}
In Eq.~\ref{eq:app:standard_objective}, $s_i(c)$ encodes the centroid index that is closest to the sample $\phi_i$. In other words, $s_i(c) = k$ if sample $\phi_i$ is closest to centroid $c_k$ in the Euclidean sense. Equation~\ref{eq:app:standard_objective} is equivalent to the standard K-means objective provided in Eq.~\ref{eq:standard_kmeans_obj} in the scalar case -- the factor of $1/2$ is inconsequential and is included above for convenience. Additionally, the function of $s_i(c)$ is equivalent to the function of the assignment matrix ${\bf L}$ in Eqs.~\ref{eq:standard_kmeans_obj} and \ref{eq:standard_indicator}. 

In the standard K-means approach, the goal is to produce a set of $K$ centroids that minimizes the above objective. To accomplish this, one path is to recover an centroid update equation from gradient descent as 
\begin{equation}
    \Delta c = \epsilon \frac{\partial E(c)}{\partial c}, 
\end{equation}
where $\epsilon$ is a so-called learning rate. Although we can proceed in the gradient descent context, a more intuitive formulation comes from the expectation-maximization (EM) perspective \cite{bengio_kmeans}, which boils down to the following question: given some previous value of the centroids, what are the values of $s_i(w)$ that minimize the objective? The derivations below proceed in the context of this question. 

\subsection{Standard K-means}
\label{sec:app:standard_kmeans}
We can cast the above question in the following cost function, which is related to the original objective defined in Eq.~\ref{eq:app:standard_objective}:
\begin{equation}
    \label{eq:app:em_objective}
    Q(c, c') = \sum_i \frac{1}{2}\norm{\phi_i - c'_{s_i(c)}}^2_2 = \sum_i \frac{1}{2} (\phi_i - c'_{s_i(c)})^2.
\end{equation}
In Eq.~\ref{eq:app:em_objective}, the current set of centroids is $c$ and the next (new) set of centroids is $c'$. The goal is to find an update rule for $c'_k$ -- the k-th centroid at the next iteration -- that minimizes the objective/cost function in Eq.~\ref{eq:app:em_objective}. This is referred to as "standard" K-means because the objective function in Eq.~\ref{eq:app:em_objective} utilizes the usual Euclidean distance. 

The analytic solution to the minimization comes from solving the following algebraic equation:
\begin{equation}
    \label{eq:app:dq_dw}
    \frac{\partial Q(c, c')}{\partial c'_k} = 0. 
\end{equation}
Because K-means is a hard clustering method, there are two conditions: $s_i(c)=k$ and $s_i(c) \neq k$. If $s_i(c)=k$, the partial derivative in Eq.~\ref{eq:app:dq_dw} becomes
\begin{equation}
\begin{aligned}
    \frac{\partial Q(c, c')}{\partial c'_k} 
    &= \frac{\partial}{\partial c'_k}\left[ \sum_i \frac{1}{2} (\phi_i - c'_k)^2 \right]\\ 
    &= \sum_i \frac{\partial}{\partial c'_k} \left[ \frac{1}{2} (\phi_i - c'_k)^2 \right]\\
    &= \sum_i c'_k - \phi_i. 
\end{aligned}
\end{equation}
If $s_i(w) \neq k$, the partial derivative in Eq.~\ref{eq:app:dq_dw} is zero. Combining these conditions leads to the expression
\begin{equation}
    \frac{\partial Q(c, c')}{\partial c'_k} = 
    \sum_i
    \begin{cases}
        c'_k - \phi_i     & \text{if } s_i(c) = k,\\
        0              & \text{otherwise.}
    \end{cases}
    \rightarrow \frac{\partial Q(c, c')}{\partial c'_k} = \sum_{i: s_i(c)=k} (c'_k - \phi_i). 
\end{equation}
Solving for Eq~\ref{eq:app:dq_dw} results in 
\begin{equation}
    \begin{aligned}
    \frac{\partial Q(c, c')}{\partial c'_k} 
    = \sum_{i: s_i(c)=k} (c'_k - \phi_i) = 0, \\
    \sum_{i: s_i(c)=k} c'_k = \sum_{i: s_i(c)=k} \phi_i,
    \end{aligned}
\end{equation}
which ultimately provides the familiar centroid update rule:
\begin{equation}
    \label{eq:app:standard_update} 
    {c'_k = \frac{1}{N_k} \sum_{i: s_i(c)=k} \phi_i, \text{ where } N_k = \sum_{i: s_i(c)=k} 1}. 
\end{equation}
In other words, the centroid at the next iteration $c'_k$ is the within-cluster sample mean using labels from the previous iteration -- convergence of the iterative procedure minimizes the target objective in Eq.~\ref{eq:app:em_objective}, which was the starting point. This is the update rule used in both Alg.~\ref{alg:standard_kmeans} and Alg.~\ref{alg:jacobian_kmeans}. 

\subsection{Constant Scaling Factor}
\label{sec:app:constant_scaled_kmeans}
Here, the standard K-means objective is modified slightly with a constant-valued, linear scaling factor $a$: 
\begin{equation}
    Q(c, c') = \sum_i \frac{1}{2}\norm{a(\phi_i - c'_{s_i(c)})}^2_2 = \sum_i \frac{1}{2} (a \phi_i - a c'_{s_i(c)})^2.
\end{equation}
Constant-valued here means $a$ is set once a-priori and fixed -- since both data points $\phi_i$ and centroids $c_k$ are scalars, $a$ in this case is also a scalar. In the case of $s_i(c) = k$, the partial derivative becomes 
\begin{equation}
    \label{eq:app:fixed_kernel_derivative}
    \begin{aligned}
    \frac{\partial Q(c, c')}{\partial c'_k} 
    &= \frac{\partial}{\partial c'_k}\left[ \sum_i \frac{1}{2} (a \phi_i - a c'_k)^2 \right]\\ 
    &= \sum_i \frac{\partial}{\partial c'_k} \left[ \frac{1}{2} (a \phi_i - a c'_k)^2 \right]\\
    &= \sum_i a^2 (c'_k - \phi_i).
    \end{aligned}
\end{equation}
Again, note that if $s_i(c) \neq k$, the partial derivative $\frac{\partial Q(c, c')}{\partial c'_k}  =0 $. This leads to 
\begin{equation}
    \label{eq:app:constant_linear} 
    \frac{\partial Q(c, c')}{\partial c'_k} = 
    \sum_i
    \begin{cases}
        a^2(c'_k - \phi_i)     & \text{if } s_i(c) = k,\\
        0              & \text{otherwise.}
    \end{cases}
    \rightarrow \frac{\partial Q(c, c')}{\partial c'_k} = \sum_{i: s_i(c)=k} a^2(c'_k - \phi_i). 
\end{equation}
Because the parameter $a$ is nonzero, solving $\frac{\partial Q(c, c')}{\partial c'_k}  =0 $ for $c'_k$ produces the same update rule as the standard case (Eq.~\ref{eq:app:standard_update}).

\subsection{Centroid-Dependent Scaling Factor}
\label{sec:app:jacobian_kmeans}
The standard K-means objective is modified here with a \textit{centroid-dependent} scaling factor $A(c_k) \in \mathbb{R}$: 
\begin{equation}
    \label{eq:app:loss_kernel}
    Q(c, c') = \sum_i \frac{1}{2}\norm{A(c'_{s_i(c)})\left(\phi_i - c'_{s_i(c)}\right)}^2_2 = \sum_i \frac{1}{2} \left( A(c'_{s_i(c)}) \phi_i -  A(c'_{s_i(c)}) c'_{s_i(c)} \right)^2.
\end{equation}
The scaling factor $A_k = A(c'_k)$ is evaluated at a specific centroid location and is directly analogous to the role played by ${\bf A}_k$ in the Jacobian-scaled K-means formulation used in Sec.~\ref{sec:jskmeans}. This means that the scaling factor $A_k$ dynamically adapts to the movement of the centroids during the K-Means iterations. As alluded in Sec.~\ref{sec:jacobian_algorithm}, the derivations below show that the standard centroid update incurs an error when augmenting the distance function in the objective with a centroid-dependent linear scaling, as in Eq.~\ref{eq:app:loss_kernel}. 

In the case of $s_i(c)=k$, the partial derivative becomes 
\begin{equation}
    \label{eq:app:kernel_derivative} 
    \begin{aligned}
    \frac{\partial Q(c, c')}{\partial c'_k} 
    &= \frac{\partial}{\partial c'_k}\left[ \sum_i \frac{1}{2} \left( A(c'_{k}) \phi_i -  A(c'_{k}) c'_{k} \right)^2 \right]\\ 
    &= \sum_i \frac{\partial}{\partial c'_k} \left[\frac{1}{2} \left( A(c'_{k}) \phi_i -  A(c'_{k}) c'_{k} \right)^2 \right]\\
    &= \sum_i \left( A(c'_{k}) \phi_i -  A(c'_{k}) c'_{k} \right) \underbrace{\frac{\partial}{\partial c'_k} \left[ A(c'_{k}) \phi_i -  A(c'_{k}) c'_{s_i(c)} \right]}_{\Psi}
    \end{aligned}
\end{equation}
Expanding for the term denoted $\Psi$ in the above equation yields
\begin{equation}
    \label{eq:app:phi}
    \begin{aligned}
    \Psi
    &= {\frac{\partial}{\partial c'_k} \left[ A(c'_{k}) \phi_i -  A(c'_{k}) c'_{s_i(c)} \right]}\\
    &= \frac{\partial}{\partial c'_k} \left[ A(c'_k) \phi_i \right] - \frac{\partial}{\partial c'_k}\left[ A(c'_k) c'_k \right]\\
    &= \phi_i \frac{\partial A(c'_k)}{\partial c'_k} - \left( c'_k \frac{\partial A(c'_k)}{\partial c'_k}  + A(c'_k) \right)\\
    &= \frac{\partial A(c'_k)}{\partial c'_k} \left ( \phi_i - c'_k \right) - A(c'_k).
    \end{aligned}
\end{equation}
Plugging back in to Eq.~\ref{eq:app:kernel_derivative}: 
\begin{equation}
    \label{eq:app:adaptive_kernel_derivative}
    \begin{aligned}
    \frac{\partial Q(c, c')}{\partial c'_k} 
    &= \sum_i \left( A(c'_{k}) \phi_i -  A(c'_{k}) c'_{k} \right) \underbrace{ \left( \frac{\partial A(c'_k)}{\partial c'_k} \left ( \phi_i - c'_k \right) - A(c'_k) \right) }_{\Psi}\\
    &= \sum_i -A(c'_k) \left( A(c'_k) \phi_i - A(c'_k) c'_k \right) + 
            \frac{\partial A(c'_k)}{\partial c'_k} (\phi_i - c'_k) \left( A(c'_k) \phi_i - A(c'_k) c'_k \right)\\
    &= \sum_i \underbrace{\frac{\partial A(c'_k)}{\partial c'_k} A(c'_k) (\phi_i - c'_k)^2}_{\text{Variance contribution}} 
            - 
              \underbrace{A(c'_k)^2 (\phi_i - c'_k)}_{\text{Fluctuation contribution}}.
    \end{aligned}
\end{equation}
Note that in Eq.~\ref{eq:app:adaptive_kernel_derivative}, the partial derivative contains the same fluctuation contribution as Eq.~\ref{eq:app:fixed_kernel_derivative}, the constant scaling case. What is introduced by the additional requirement of centroid-dependency in the scaling factor, however, is an additional contribution that (a) depends on a gradient of $A(c'_k)$, and (b) scales with the within cluster variance expressed as $(\phi_i - c'_k)^2$. This modification can also be interpreted as the inclusion of a higher-order term in the form of an additional within-cluster moment. 

Recall again that in the case of $s_i(c) \neq k$, we get the usual $\frac{\partial Q(c, c')}{\partial c'_k} = 0$. As such, an iterative solution to the minimization problem requires solving the following equation for $c'_k$: 
\begin{equation}
    \label{eq:app:variance_contrib}
    {\frac{\partial Q(c, c')}{\partial c'_k} 
    = \underbrace{ \sum_{i: s_i(c)=k} \frac{\partial A(c'_k)}{\partial c'_k} A(c'_k) (\phi_i - c'_k)^2}_{\text{Variance contribution}} 
            - 
              \underbrace{ \sum_{i: s_i(c)=k} A(c'_k)^2 (\phi_i - c'_k)}_{\text{Fluctuation contribution}} = 0.}
\end{equation}
The key takeaway is that if the variance contribution diminishes, the update rule for the centroids becomes the standard update of Eq.~\ref{eq:app:standard_update}. The above equation can therefore be interpreted as deviation from the standard centroid update rule by a residual proportional to the variance contribution. As implied by Eq.~\ref{eq:app:variance_contrib}, a variance contribution of zero would require one the following conditions to be met: 
\begin{itemize}
    \item The gradient of the scaling factor within the cluster is zero. 
    \item The within-cluster variance is zero. 
    \item The scaling factor itself is zero.
\end{itemize}
If the scale factor $A(c_k)$ is derived from the Jacobian of a chemical source term (as demonstrated in the main text), the first condition cannot be valid in regions of high chemical sensitivity. Further, within this context, the second and third conditions are met only in trivial equilibrium and ambient conditions in which there is no chemical contribution to the dynamics (i.e. zero source term). As such, a centroid update that does not take into account the variance contribution in Eq.~\ref{eq:app:variance_contrib} cannot guarantee monotonic convergence of the modified objective function in Eq.~\ref{eq:app:loss_kernel} -- this is consistent with the non-monotonic convergence trends observed for the Jacobian-scaled K-means algorithm in Sec.~\ref{sec:detonation_results}.

\end{document}